\documentclass[12pt,english]{article}
\usepackage[utf8]{inputenc}

\usepackage[left=1in,right=1in,top=1in,bottom=1in]{geometry}

\usepackage{graphicx}
\usepackage{subfigure}
\usepackage{braket}
\usepackage{amsmath}
\usepackage{amssymb}

\usepackage{physics}
\usepackage{leftidx}
\usepackage{feynmf}
\usepackage{wrapfig} 

\usepackage{prettyref}

\usepackage{esint}

\usepackage{dsfont}

\usepackage{color}
\definecolor{darkgreen}{rgb}{0,0.5,0}
\definecolor{darkblue}{rgb}{0,0,0.6}
\definecolor{purple}{rgb}{0.4,.2,0.7}
\usepackage[colorlinks=true,citecolor=darkgreen,linkcolor=purple,urlcolor=purple]{hyperref}

\usepackage{enumitem}
\usepackage{ulem} 
\usepackage{tensor} 
\usepackage{slashed}
\usepackage{mathtools}

\numberwithin{equation}{section}
\numberwithin{figure}{section}
\numberwithin{table}{section}

\begin{document}

\begin{center}
{\Large A Compendium of Sphere Path Integrals}  
\end{center}

\vskip1cm

\begin{center}
  Y.T.\ Albert Law
\vskip5mm
{\it{\footnotesize  Department of Physics, Columbia University, New York, NY 10027}
}
\vskip5mm
{\footnotesize  Email: \href{mailto:yal2109@columbia.edu}{yal2109@columbia.edu}}
\end{center}

\vskip2cm

\vskip0mm
\noindent {\bf Abstract:} We study the manifestly covariant and local 1-loop path integrals on $S^{d+1}$ for general massive, shift-symmetric and (partially) massless  totally symmetric tensor fields of arbitrary spin $s\geq 0$ in any dimensions $d\geq 2$. After reviewing the cases of massless fields with spin $s=1,2$, we provide a detailed derivation for path integrals of massless fields of arbitrary integer spins $s\geq 1$. Following the standard procedure of Wick-rotating the negative conformal modes, we find a higher spin analog of Polchinski's phase for any integer spin $s\geq 2$. The derivations for low-spin ($s=0,1,2$) massive, shift-symmetric and partially massless fields are also carried out explicitly. Finally, we provide general prescriptions for general massive and shift-symmetric fields of arbitrary integer spins and partially massless fields of arbitrary integer spins and depths.


\newpage

\tableofcontents



\section{Introduction}

Arising as the leading saddle point for the gravitational Euclidean path integral with a positive cosmological constant, the sphere plays a dominant role in the study of quantum gravity in de Sitter space \cite{PhysRevD.15.2752, Gibbons:1978ji, Christensen:1979iy,Fradkin:1983mq, Allen:1983dg,Taylor:1989ua,GRIFFIN1989295, Mazur:1989ch, Vassilevich:1992rk, Volkov:2000ih, Anninos:2020hfj,Polchinski:1988ua}. From the static patch point of view, quadratic fluctuations around the sphere saddle contribute to the 1-loop corrections to the Gibbons-Hawking de Sitter horizon entropy. Recently, envisioning a 1-loop test for microscopic models for de Sitter horizon, the authors in \cite{Anninos:2020hfj} derived an integral formula for 1-loop sphere path integrals in terms of Harish-Chandra characters for the de Sitter group.

For example, the character formula for a scalar field with generic mass $m^2$ on $S^{4}$ (with radius set to 1) is \cite{Anninos:2020hfj}
\begin{align}\label{massive char}
	\log Z_\text{PI}=\int_0^\infty \frac{dt}{2t}\frac{1+e^{-t}}{1-e^{-t}} \,\chi_\Delta(t)\,,
\end{align}
where 
\begin{align}\label{Characters}
	\chi_\Delta(t)=\frac{e^{-\Delta t}+e^{-\bar{\Delta} t}}{(1-e^{-t})^3}
\end{align}
is the Harish-Chandra character of the isomatry group $SO(1,4)$ of $dS_{4}$. The scaling dimension $\Delta$ is related to the mass $m^2$ through $m^2=(\Delta-2)(\bar{\Delta}-2)$ and $\bar{\Delta}=3-\Delta$. As explained in \cite{Anninos:2020hfj}, $\chi(t)$ captures massive scalar quasinormal modes in a de Sitter static patch. With the formula \eqref{massive char} we can compute exact 1-loop contribution by a massive scalar to de Sitter horizon entropy.

The present work goes back to the starting point of the derivation of the character formula, that is, the 1-loop sphere path integral itself
\begin{align}\label{PI general}
	Z_\text{PI} =\int \mathcal{D}\phi \, e^{-S[\phi]}
\end{align}
where $S[\phi]$ is the quadratic action of the field $\phi$. Typically, such an object is expressed in terms of functional determinants of kinetic operators. A massive real scalar has
\begin{align}\label{intro scalar det}
	Z_\text{PI}=\det\left(-\nabla^2+m^2\right)^{-1/2}
\end{align}
which can be massaged into the formula \eqref{massive char} as shown in \cite{Anninos:2020hfj}. However, the computation becomes more subtle and intricate for fields with spin $s\geq 1$. For instance, it took decades of work \cite{PhysRevD.15.2752, Gibbons:1978ji, Christensen:1979iy,Fradkin:1983mq, Allen:1983dg,Taylor:1989ua,GRIFFIN1989295, Mazur:1989ch, Vassilevich:1992rk, Volkov:2000ih,Polchinski:1988ua} before the correct generalization of \eqref{intro scalar det} was obtained for a massless spin-2 field on $S^4$. 

To generalize the formula \eqref{massive char} for wider classes of field contents, one must first obtain their correct functional determinant expressions. This is the main motivation for this work. In this paper we perform detailed derivations starting from manifestly local covariant path integrals on $S^{d+1}$. Our goals are twofold: 1. address subtlties of 1-loop sphere path integrals and clarify previous computations; 2. generalize all known results to any massive, shift-symmetric and (partially) massless  totally symmetric tensor fields of arbitrary spin $s\geq 0$ in any dimensions $d\geq 2$. 

As an illustration, starting from \eqref{PI general} with the free massless spin-$s$ Fronsdal action, a lengthy derivtion in section \ref{Massless PI} leads to the following expression (equation \eqref{HS final}) for massless higher-spin theories
\begin{align}\label{intro HS PI}
	Z^\text{HS}_\text{PI}=
	i^{-P}
	\frac{\gamma^{\text{dim}\,G}}{\text{Vol($G$)}_\text{can}} \prod_s  \left( (d+2s-2)(d+2s-4)\right)^{\frac{N^\text{KT}_{s-1}}{2}}
	\frac{\det\nolimits'_{-1}\left|-\nabla_{(s-1)}^2-\lambda_{s-1,s-1}\right|^{1/2}}{ \det\nolimits'_{-1}\left|-\nabla_{(s)}^2-\lambda_{s-2,s}\right|^{1/2}} 
\end{align}
Here we highlight a few features of this expression:
\begin{itemize}
	\item In the ratio of ghost and physical functional determinants, $-\nabla_{(s)}^2$ is the spin-$s$ symmetric transverse traceless (STT) Laplacians on $S^{d+1}$ and $\lambda_{n,s}$ are its eigenvalues. Their relevant properties are summarized in appendix \ref{STSH}.  The prime denotes omission of zero modes of the Laplace operators. The subscript -1 is related to the contribution from longitudinal modes, which was obtained in \cite{Anninos:2020hfj} by demanding the absence of logarithmic divergence for odd $d+1$. Here we obtain these by a direct path integral computation.
	
	\item The second factor is associated with the group $G$ of trivial gauge transformations. $\gamma$ is related to the coupling constant of the theory, while $\text{Vol}(G)_\text{can}$ is what was called canonical group volume in \cite{Anninos:2020hfj}. It was emphasized in \cite{Donnelly:2013tia} that the inclusion of this factor was crucial for consistency with locality and unitarity. The peculiar factor $\prod_s  \left((d+2s-2)(d+2s-4)\right)^{\frac{N^\text{KT}_{s-1}}{2}}$ arises when we relate the metric on the space of trivial gauge transformations induced by the path integral measure to the canonical metric to be defined precisely below.
	
	\item  The phase factor $i^{-P}$ is present only for fields with spin $s\geq 2$, whose origin is the negative conformal modes that render the Euclidean path integral divergent. The standard prescription \cite{Gibbons:1978ac} is to Wick rotate the problematic conformal modes in field space so that the integrals converge. Polchinski later \cite{Polchinski:1988ua} found that on $S^{d+1}$ this procedure led to a finite number of $i$ factors (with $P=d+3$ in that case) that could leave the Euclidean path integral positive, negative or imaginary depending on the dimensions. 
\end{itemize}
Analogous expressions are derived for any massive (equation \eqref{massive general}), shift-symmetric (equation \eqref{shift general}) and partially massless (equation \eqref{general s t}) totally symmetric tensor fields of arbitrary spin $s\geq 0$ in any dimensions $d\geq 2$. With these precise expressions one could then derive a formula analogous to \eqref{massive char} for more general representations. As for most quantities in quantum field theory, objects such as \eqref{intro HS PI} are UV-divergent. The recipe for exactly evaluating these formal expressions and their UV divergences were discussed in \cite{Anninos:2020hfj}.

Finally, although this work is primarily motivated by the study of de Sitter thermodynamics, sphere partition functions are of interest in a broad range of contexts such as string theory, Chern-Simons theory, supersymmetry, AdS/CFT correspondence, conformal field theory, as well as entanglement entropy in quantum field theories. We anticipate that the 1-loop results contained in this paper will be relevant in these contexts as well.

\paragraph{Plan of the paper:} We first review the computations for massless spin-1 and spin-2 fields in sections \ref{massless spin1} and \ref{massless spin2}. We then turn to our complete derivation for massless fields of arbitrary integer spins in section \ref{Massless PI}. In section \ref{Massive PI}, we study fields with generic mass. In sections \ref{shift sym} and \ref{PM fields}, we study general shift-symmetric fields and partially massless fields respectively. We conclude in section \ref{conclusion}. All conventions are summarized in appendix \ref{convention}. Relevant properties of the STT Laplacians on $S^{d+1}$ and their eigenfunctions are collected in appendix \ref{STSH}. The higher spin invariant bilinear form is reviewed in appendix \ref{cubic}.


\section{Review of massless vectors}\label{massless spin1}

We start with a pedagogical review of the case of massless vectors. The object of interest is the 1-loop approximation to the full Euclidean path integral 
\begin{align}\label{full vector PI}
	Z_\text{PI} =\frac{1}{\text{Vol}(\mathcal{G})}\int \mathcal{D} A^a \mathcal{D}\Phi \, e^{-S_E[A^a,\Phi]}
\end{align}
for a theory that involves a collection of massless vector (for example U(1) or Yang-Mills) gauge fields interacting with some matter fields, denoted as $A\indices{^a_\mu}$ and collectively as $\Phi$ respectively, living on $S^{d+1}$. 


\paragraph{$U(1)$ with a complex scalar}

The simplest example involves a single U(1) gauge field $A_\mu$ interacting with a complex scalar $\phi$ (studied in \cite{Allen:1983dg}): 
\begin{align}\label{U1 ex}
	S_E[A, \phi]=\int_{S^{d+1}} \bigg[ \frac{1}{4\mathrm{g}^2} F_{\mu\nu}F^{\mu\nu}+ D_{\mu}\phi (D^\mu \phi)^* +m^2 \phi \phi^*\bigg],
\end{align}
where 
\begin{align}
	F_{\mu\nu}\equiv\partial_\mu A_\nu-\partial_\nu A_\mu,\qquad D_\mu \phi \equiv (\partial_\mu -i A_\mu)\phi
\end{align}
are the field strength and the covariant derivative of the scalar. This action is invariant under the local U(1) gauge transformations
\begin{align}
	\phi(x) \to e^{i\alpha (x)}\phi(x),\quad A_\mu(x)\to A_\mu(x)+\partial_\mu \alpha(x).
\end{align}
The normalization adopted here is to emphasize the presence of the coupling constant $\mathrm{g}$. In this convention $\mathrm{g}$ does not show up in the gauge transformation. 


\paragraph{Yang-Mills}

Another example is Yang-Mills (YM) theory with a Lie algebra
\begin{align}\label{Lie al}
	[L^a,L^b]=f^{abc}L^c
\end{align}
generated by some standard basis of anti-hermitian matrices and $f^{abc}$ is real and totally antisymmetric. The YM action is 
\begin{align}\label{YM ex}
	S_E[A, \phi]= \frac{1}{4\mathrm{g}^2}\int_{S^{d+1}} \text{Tr}F^2= \frac{1}{4\mathrm{g}^2}\int_{S^{d+1}} F_{\mu\nu}^a F^{a, \mu\nu},
\end{align}
where the curvature is $F_{\mu\nu}\equiv\partial_\mu A_\nu-\partial_\nu A_\mu+[A_\mu,A_\nu]$ with $A_\mu=A_\mu^a L^a$. Here the overall normalization for the trace (or Killing form) is {\it defined} such that the generators $L^a$ are unit normalized: 
\begin{align}\label{trace def}
	\text{Tr}(L^a L^b)\equiv\delta^{ab}.
\end{align} 
For SU(2) YM, $L^a=-\frac{i \sigma^a}{2}$ satisfying $[L^a,L^b]=\epsilon^{abc}L^c$, and the trace \eqref{trace def} would be $ \text{Tr}\equiv -2 \text{tr}$ with tr being the matrix trace. The YM action is invariant under the non-linear gauge transformations $\alpha=\alpha^a L^a$
\begin{align}
	A_\mu \to A_\mu+\partial_\mu \alpha+ [A_\mu,\alpha].
\end{align}

In both the $U(1)$ and YM examples, the corresponding path integral in \eqref{full vector PI} clearly overcounts gauge equivalent configurations. A factor $\text{Vol}(\mathcal{G})$ is thus inserted in \eqref{full vector PI} to quotient out configurations connected by gauge transformations. This factor is formally the volume of the space of gauge transformations $\mathcal{G}$ (the measure with respect to which the volume is defined will be discussed in later subsection) and is theory dependent. For the U(1) example, $\text{Vol}(\mathcal{G})$ is simply a path integral over a single local scalar field
\begin{align}
	\text{Vol}(\mathcal{G})_\text{U(1)}=\int \mathcal{D}\alpha,
\end{align}
while for SU(2) YM it would be a path integral over $3$ local scalar fields
\begin{align}
	\text{Vol}(\mathcal{G})_\text{SU(2)}=\int \mathcal{D}\alpha_1\mathcal{D}\alpha_2\mathcal{D}\alpha_3.
\end{align}
More generally, $\text{Vol}(\mathcal{G})$ is an integral over $N=\text{dim} \,G$ local scalar fields for a gauge group $G$.

\paragraph{1-loop approximation} 

Now, suppose the equation of motion admits the trivial solution $A\indices{^a_\mu}=0=\Phi$, around which we perform a saddle point approximation for \eqref{full vector PI}. Then at the quadratic (1 loop) level the vector and matter fields decouple:
\begin{gather}
	Z^\text{1-loop}_\text{PI} =Z^{\delta A}_\text{PI} Z^{\delta \Phi}_\text{PI}.
\end{gather}
In the following, we focus on the vector part of the 1-loop path integral (with $A^a$ understood as the fluctuations around the background)
\begin{align}\label{spin 1 1loop}
	Z^{\delta A}_\text{PI}=\frac{1}{\text{Vol}(\mathcal{G})}\int \prod_{a=1}^{\text{dim}\,G} \mathcal{D} A^a \, e^{-\sum_{a=1}^{\text{dim}\,G} S_E[A^a]}.
\end{align}
where $S_E[A^a]$ is simply a Maxwell action 
\begin{align}\label{Maxwell action}
	S_E[A^a]= \frac{1}{4\mathrm{g}^2} \int_{S^{d+1}} F^a_{\mu\nu}F^{a, \mu\nu}, \quad F_{\mu\nu} =\partial_\mu A^a_\nu-\partial_\nu A^a_\mu\, .
\end{align}
A careful analysis of the Euclidean path integral for the U(1) theory on arbitrary manifolds has been presented in \cite{Donnelly:2013tia}, where the authors point out the importance of taking care of zero modes, large gauge transformations and non-trivial bundles for consistency with locality and unitarity. 

In the following we will express $Z^0_\text{PI}$ in terms of functional determinants and highlight the relevant subtleties in our case of $S^{d+1}$ along the way. Before doing so we would like to make a last comment. Even though the problem is largely reduced to that of a free Maxwell theory on $S^{d+1}$, the knowledge of the parent theory \eqref{full vector PI} to which we performed the 1-loop approximation is required for the full determination of the \eqref{spin 1 1loop}. In particular, the group volume factor $\text{Vol}(\mathcal{G})$ will eventually lead to a residual group volume factor in the final result, whose value depends on the specific gauge group (e.g. $U(1)$ or $SU(N)$) and couplings of the gauge field to the matter fields.

\subsection{Transverse vector determinant and Jacobian}

\subsubsection*{Geometric approach and change of variables}

Since the path integrations over $A^a$ in \eqref{spin 1 1loop} are decoupled, we can focus on one of the factors, and we will suppress the index $a$. Traditional ways to proceed include Faddeev-Popov or BRST gauge fixing (as done in \cite{Donnelly:2013tia} for example). Here instead we take the ``geometric approach'' \cite{Babelon:1979wd,Mazur:1989by,Bern:1990bh}, which manifests its advantages when we deal with massless higher spin fields later. In this approach one changes the field variables by decomposing
\begin{align}\label{spin 1 CoV}
	A_\mu = A^T_\mu+\partial_\mu \chi 
\end{align}
where $A^T_\mu$ is the transverse or on-shell part of $A_\mu$ satisfying $\nabla^\mu A^T_\mu=0$, and $\chi$ is the longitudinal or pure gauge part of $A_\mu$. Since $S^{d+1}$ is compact, the scalar Laplacian has a normalizable constant $(0,0)$ mode, which must be excluded from the path integration for the change of variables \eqref{spin 1 CoV} to be unique
\begin{align}
	\mathcal{D}A=J\,\mathcal{D}A^T\mathcal{D}' \chi 
\end{align}
where prime denotes the exclusion of the $(0,0)$ mode. We will find the Jacobian $J$  for the change of variables \eqref{spin 1 CoV} below.

\subsubsection*{Action for $A^T_\mu$}

Because the gauge invariance of the action, $\chi$ simply drops out upon substituting \eqref{spin 1 CoV}
\begin{align}
	S_E[A^T,\chi]=\frac{1}{2\mathrm{g}^2} \int_{S^{d+1}} [A^T_\mu (-\nabla_{(1)}^2 +d) A_T^\mu ]
\end{align}
where $-\nabla_{(1)}^2$ is the trasnverse Laplacian on $S^{d+1}$. Now we expand $A^T_\mu$ in terms of spin-1 transverse spherical harmonics (see appendix \ref{STSH} for their basic properties):
\begin{align}
	A^T_\mu = \sum_{n=1}^\infty c_{n,1} f_{n,\mu}
\end{align}
and the integration measure in our convention is
\begin{align}
	\mathcal{D}A^T = \prod_{n=1}^\infty \frac{dc_{n,1}}{\sqrt{2\pi}\mathrm{g}}.
\end{align}
Performing the path integration over these modes we have
\begin{align}
	\int \mathcal{D} A e^{-S_E[A]}= J \, \det(-\nabla_{(1)}^2 +d)^{-1/2} \int \mathcal{D}' \chi 
\end{align}

\subsubsection*{Jacobian} 

We find the Jacobian $J$ by requiring consistency with the normalization condition
\begin{align}\label{spin 1 norm}
	1=\int \mathcal{D}A \, e^{-\frac{1}{2\mathrm{g}^2} (A,A)}=\int J \,\mathcal{D}A^T\mathcal{D}' \chi  \, e^{-\frac{1}{2\mathrm{g}^2} (A^T+\nabla \chi,A^T+\nabla \chi)}.
\end{align}
Since $A^T$ is transverse, we have
\begin{align}
	(A^T+\nabla \chi,A^T+\nabla \chi)= (A^T,A^T)+ (\nabla \chi,\nabla \chi).
\end{align}
We can then path integrate $A^T$ trivially. We expand $\chi$ in terms of scalar spherical harmonics:
\begin{align}
	\chi = \sum_{n=1}^\infty c_{n,0} f_n
\end{align}
with path integration measure
\begin{align}
	\mathcal{D}'\chi = \prod_{n=1}^\infty \frac{dc_{n,0}}{\sqrt{2\pi}\mathrm{g}}.
\end{align}
Plugging this into \eqref{spin 1 norm} results in 
\begin{align}\label{vector Jac}
	J = \det\nolimits'(-\nabla_{(0)}^2)^{1/2},
\end{align}
where the prime denotes the omission of the constant $(0,0)$ mode.

\subsection{Residual group volume}\label{s 1 vol}


Let us go back to the full 1-loop path integral \eqref{spin 1 1loop}. So far we have 
\begin{align}
	Z^{\delta A}_\text{PI}=\frac{\int \prod_{a=1}^{\text{dim} G} \mathcal{D}'\chi^a}{\text{Vol}(\mathcal{G})} \left(\frac{ \det'(-\nabla_{(0)}^2)^{1/2}}{\det(-\nabla_{(1)}^2 +d)^{1/2}} \right)^{\text{dim} G}
\end{align}
where we have restored the color index $a$. Now we focus on the factor
\begin{align}
	\frac{\int \prod_{a=1}^{\text{dim} G} \mathcal{D}'\chi^a}{\text{Vol}(\mathcal{G})}.
\end{align}
As explained above, the factor $\text{Vol}(\mathcal{G})$ is theory dependent and is formally an integral over $N={\text{dim} G}$ \textbf{local} scalar fields 
\begin{align}
	\text{Vol}(\mathcal{G}) =\int \prod_{n=1}^{\text{dim} G} \mathcal{D}\alpha_n .
\end{align}
In particular, the integral includes integrations over constant scalar modes. As explained in \cite{Donnelly:2013tia}, the inclusion of zero modes is crucial for consistency with locality and unitarity. Thus, this factor does not cancel completely with the integrations over $\chi$, leaving a factor
\begin{align}\label{vec group fac}
	\frac{\int \prod_{a=1}^{\text{dim} G} \mathcal{D}'\chi^a}{\text{Vol}(\mathcal{G})}=\frac{1}{\text{Vol}(G)_\text{PI}},\quad \text{Vol}(G)_\text{PI} \equiv \int \prod_{a=1}^{\text{dim}\,G} \frac{d\alpha^a_0}{\sqrt{2\pi}\mathrm{g}}.
\end{align}
where $\alpha^a_0$ is the expansion coefficient of the $(0,0)$ mode of $\alpha^a$ ($a$ is the color index)
\begin{align}
	\alpha^a = \sum_{n=0}^\infty \alpha^a_{n} f_n.
\end{align}
These constant scalar modes correspond to the gauge transformations that leave the background $A^\mu=0$ invariant, or equivalently whose linear part is trivial. If the original full theory contains matter fields such as \eqref{U1 ex}, these act non-trivially on the latter. $G$ is therefore the group of \textbf{global} symmetries of the theory and $\text{Vol}(G)_\text{PI}$ is the volume of $G$. Note that the precise value of $\text{Vol}(G)_\text{PI}$ depends on the metric on $G$. We have been using a specific choice of metric
\begin{align}\label{YM PI metric}
	ds_\text{PI}^2  = \frac{1}{2\pi \mathrm{g}^2} \int_{S^{d+1}} \text{Tr}(\delta \alpha  \delta \alpha)
\end{align}
induced by our convention for the path integral measure. Note that had we normalized the generators $L^a$ in a different way: $L^a \to \lambda L^a$  (or equivalently choosing a  different overall normalization for the trace in the action \eqref{YM ex}: $\text{Tr} \to \lambda^2 \text{Tr}$), the path integral describes the same physics if we rescale $\mathrm{g}\to \lambda \mathrm{g}$. In particular, the metric \eqref{YM PI metric} remains the same. We want to relate the volume $\text{Vol}(G)_\text{PI}$ measured in this metric to a ``canonical volume'' $\text{Vol}(G)_\text{can}$, defined as follows. A general group element in $G$ takes the form
\begin{align}
	e^{\theta \cdot \hat{L}}=e^{\theta^a \hat{L}^a}
\end{align}
where $\hat{L}^a$ are unit-normalized. We define $\text{Vol}(G)_\text{can}$ to be the volume of the space spanned by $\theta$. In our convention, $L^a$ are unit-normalized, and therefore the relation between the metric \eqref{YM PI metric} (restricted to the subspace of trivial gauge transformations) and the canonical metric is simply
\begin{align}\label{pi metric can}
	ds_\text{PI}^2  = \frac{1}{2\pi \mathrm{g}^2} \sum_{a}(d\alpha_0^a)^2= \frac{1}{2\pi \mathrm{g}^2} \sum_{a}\left(\frac{d\theta^a}{f_0}\right)^2= \frac{\text{Vol}(S^{d+1})}{2\pi \mathrm{g}^2} ds_\text{can}^2,\quad ds_\text{can}^2 \equiv d\theta \cdot d\theta.
\end{align}
Thus we can express the group volume as
\begin{align}\label{PI volume can}
	\text{Vol}(G)_\text{PI} = \left( \frac{\text{Vol}(S^{d+1})}{2\pi \mathrm{g}^2}\right)^\frac{\text{dim}(G)}{2}\text{Vol}(G)_\text{can} = \bigg( \frac{\text{Vol}(S^{d-1})}{d \mathrm{g}^2}\bigg)^\frac{\text{dim}(G)}{2}\text{Vol}(G)_\text{can},
\end{align}
where we have used $\text{Vol}(S^{d+1})=\frac{2\pi}{d}\text{Vol}(S^{d-1})$ in the last step. The canonical volume $\text{Vol}(G)_\text{can}$ so defined is evidently independent of the coupling. To summarize, the full 1-loop path integral is
\begin{align}
	Z^{\delta A}_\text{PI}=&Z_\text{G}Z_\text{Char}\nonumber\\
	Z_\text{G}=&\frac{\gamma^{\text{dim} G}}{\text{Vol}(G)_\text{can}},\quad  \gamma=\frac{\mathrm{g}}{\sqrt{(d-2)\text{Vol}(S^{d-1})}}\nonumber\\
	Z_\text{Char}=&\left(d(d-2)\right)^{\frac{1}{2}{\text{dim} G}}\left(\frac{ \det'(-\nabla_{(0)}^2)}{\det(-\nabla_{(1)}^2 +d)}\right)^{\frac{1}{2}{\text{dim} G}}
\end{align}
The notation $Z_\text{Char}$ emphasizes that this part can be re-written in terms of the $SO(1,d+1)$ characters as explained in \cite{Anninos:2020hfj}. In retrospect, the coupling dependence of the result is precisely encoded in the group volume factor $\text{Vol}(G)_\text{PI}$. In the $G=U(1)$ example, $\text{Vol}(G)_\text{can}=\text{Vol}(U(1))_c =2\pi$, and the full 1 loop vector path integral is therefore 
\begin{align}
	Z^{U(1)}_\text{PI}=\frac{ \mathrm{g}}{ \sqrt{2\pi\text{Vol}(S^{d+1})}} \frac{ \det'(-\nabla_{(0)}^2)^{1/2}}{\det(-\nabla_{(1)}^2 +d)^{1/2}},
\end{align}
which reproduces eq.(2.6) in \cite{Giombi:2015haa}. For $G=SU(2)$, $\text{dim}\, G=3$ and  $\text{Vol}(G)_\text{can}=16\pi^2$, and thus
\begin{align}
	Z^{SU(N)}_\text{PI}=\frac{1}{16\pi^2}\bigg( \frac{2\pi \mathrm{g}^2}{ \text{Vol}(S^{d+1})}\bigg)^\frac{3}{2}\left(\frac{ \det'(-\nabla_{(0)}^2)}{\det(-\nabla_{(1)}^2 +d)} \right)^{3/2}.
\end{align}
As a concrete example for an exact evaluation of such a formal expression, with the recipe in \cite{Anninos:2020hfj} one can find for the 1-loop path integral for $SU(4)$ Yang-Mills on $S^5$ 
\begin{align} \label{MaxwellS5}
	\log Z_{\rm PI}^{SU(4), \,S^5} = 15\left( \frac{9 \pi }{8}  \frac{\ell^5}{\epsilon^5}-\frac{5 \pi }{8}\frac{\ell^3}{\epsilon^3}-\frac{7 \pi }{16} \frac{\ell}{\epsilon}\right)+ \log \frac{\left(\mathrm{g}/\sqrt{\ell}\right)^{15}}{{\frac{1}{6}(2\pi)^9}} + 15  \left( \, \frac{5 \, \zeta (3)}{16 \, \pi ^2} +\frac{3 \, \zeta (5)}{16 \, \pi ^4} \, \right)  
\end{align}
where we have restored the sphere radius $\ell$ and $\epsilon$ is the UV regulator in heat kernel regularization.


\subsubsection*{Local gauge algebra, global symmetry and invariant bilinear form}

For the later discussions on spin 2 and massless higher spin fields, and to make connection with the work in \cite{Joung:2013nma}, we offer another perspective for the non-abelian case.

\paragraph{Local gauge algebra}

Recall that the original Yang-Mills action \eqref{YM ex} is invariant under the full non-linear infinitesimal gauge transformations
\begin{gather}
	\delta_\alpha A_\mu =\delta^{(0)}_\alpha A_\mu +\delta^{(1)}_\alpha A_\mu \nonumber\\
	\delta^{(0)}_\alpha A_\mu= \partial_\mu \alpha ,\qquad \delta^{(1)}_\alpha A_\mu= [A_\mu,\alpha].
\end{gather}
Here the superscript $(n)$ denotes the power in fields. This generates an algebra
\begin{align}\label{YM local gauge algebra}
	\delta_\alpha \delta_{\alpha'}A_\mu- \delta_{\alpha'}\delta_\alpha A_\mu=\delta_{[[\alpha,\alpha']]}A_\mu
\end{align}
where we have defined a bracket $[[\cdot,\cdot ]]$ on the space of gauge parameters, which in our convention is equal to the negative of the matrix commutator\footnote{One should keep in mind that $[[\cdot,\cdot]]$ is defined using the gauge transformations of $A_\mu$, whose precise form depends on the normalization conventions, while the commutator on the right hand side is the matrix commutator $[A,B]=AB-BA$. Had we normalized $A_\mu$ canonically, so that the action takes the form $- \frac{1}{4}\int_{S^{d+1}} \text{Tr}F^2$, the gauge transformations will be instead $\delta_\alpha A_\mu =\partial_\mu \alpha+\mathrm{g} [A_\mu,\alpha]$ and the local gauge algebra will become $[\delta_\alpha, \delta_{\alpha'}]=\delta_{-\mathrm{g}[\alpha,\alpha']}$ and the bracket will read $[[\bar{\alpha},\bar{\alpha}']]=-\mathrm{g}[\alpha,\alpha']$.}
\begin{align}\label{YM bracket}
	[[\alpha,\alpha']]=-[\alpha,\alpha'].
\end{align}

\paragraph{Global symmetry algebra from the gauge algebra}

The constant $(0,0)$ modes $\bar{\alpha}$ generate background ($A_\mu=0$) preserving gauge transformations satisfying 
\begin{align}
	\delta^{(0)}_{\bar{\alpha}}=0,
\end{align}
which form a subalgebra $\mathfrak{g}$ of the local gauge algebra, with the bracket $[[\cdot,\cdot]]$ naturally inherited from the local gauge algebra
\begin{align}
	[[\bar{\alpha},\bar{\alpha}']]=-[\bar{\alpha},\bar{\alpha}']. 
\end{align}
This global symmetry algebra $\mathfrak{g}$ is clearly isomorphic to the original Lie algebra \eqref{Lie al}. On $\mathfrak{g}$, the path integral metric \eqref{YM PI metric} corresponds to the bilinear form with a specific normalization:
\begin{align}\label{vec PI bi form}
	\bra{\bar{\alpha}}\ket{\bar{\alpha}'}_\text{PI}=\frac{1}{2\pi \mathrm{g}^2} \int_{S^{d+1}} \bar{\alpha}^a \bar{\alpha}'^{ a}=\frac{\text{Vol}(S^{d+1})}{2\pi \mathrm{g}^2} \bar{\alpha}^a \bar{\alpha}'^{ a}.
\end{align}
We define a theory independent ``canonical'' invariant bilinear form $\bra{\cdot}\ket{\cdot}_\text{c}$ on $\mathfrak{g}$ as follows.
\begin{enumerate}
	\item Pick a basis $M^a$ of $\mathfrak{g}$ such that they satisfy the same commutation relation as $L^a$: $[[M^a,M^b]]=f^{abc}M^c$. This fixes the relative normalizations of $M^a$. 
	
	\item Fix the overall normalization of $\bra{\cdot}\ket{\cdot}_\text{c}$ by requiring $M^a$ to be unit-normalized:
	\begin{align}
		\bra{M^a}\ket{M^b}_\text{c}=\delta^{ab}
	\end{align}
\end{enumerate}
In the current case, this means that we should take $M^a=L^a$ and 
\begin{align}
	\bra{\alpha}\ket{\alpha'}_\text{c}= \bar{\alpha}^a \bar{\alpha}'^{ a}.
\end{align}
Comparing this with  \eqref{vec PI bi form}, we see that the path integral and canonical metrics are related as in \eqref{pi metric can}, leading to the same result \eqref{PI volume can}.


\section{Review of massless spin 2}\label{massless spin2}

Next we review the computation for linearized Einstein gravity on $S^{d+1}$, which has a long and dramatic history \cite{Gibbons:1978ji, Christensen:1979iy,Fradkin:1983mq, Allen:1983dg,Taylor:1989ua,GRIFFIN1989295, Mazur:1989ch, Vassilevich:1992rk, Volkov:2000ih,Polchinski:1988ua}. Expanding the gravitational path integral with the Einstein-Hilbert action $\frac{1}{16\pi G_N}\int_{S^{d+1}} (2\Lambda-R)$ around the $S^{d+1}$ saddle: $g_{\mu\nu}=g^{S^{d+1}}_{\mu\nu}+h_{\mu\nu}$ up to quadratic order, we obtain the Euclidean path integral 
\begin{align}\label{s2pathintegral}
	Z_\text{PI}=\frac{1}{\text{Vol($\mathcal{G}$)}}\int \mathcal{D}h \, e^{-S[h]}
\end{align}
where the action for a massless spin-2 particle on $S^{d+1}$ is 
\begin{align}\label{eq:spin2action}
	S[h] = \frac{1}{2\mathrm{g}^2}\int_{S^{d+1}}h^{\mu\nu}\bigg[ (-\nabla^2+2)h_{\mu\nu}+2 \nabla_{(\mu}\nabla^\lambda h_{\nu) \lambda}+g_{\mu\nu}(\nabla^2 h\indices{_\lambda^\lambda}-2\nabla^{\sigma}\nabla^\lambda h_{\sigma \lambda})+(D-3)g_{\mu\nu}h\indices{_\lambda^\lambda}\bigg],
\end{align}
where $\mathrm{g}=\sqrt{32\pi G_N}$. \eqref{eq:spin2action} is invariant under the linearized diffeomorphisms\footnote{The insertion of the factor $\frac{1}{\sqrt{2}}$ is for later convenience.}
\begin{align}\label{lin diff}
	h_{\mu\nu}\to h_{\mu\nu}+\sqrt{2}\nabla_{(\mu}\Lambda_{\nu)}=h_{\mu\nu}+\frac{1}{\sqrt{2}}(\nabla_{\mu} \Lambda_{\nu}+\nabla_{\nu} \Lambda_{\mu}).
\end{align}
The factor $\text{Vol($\mathcal{G}$)}$ is the volume of the space of diffeomorphisms, which is a path integral over a local vector field $\alpha_\mu$
\begin{align}\label{vol g local vec}
	\text{Vol}(\mathcal{G}) =\int \mathcal{D}\alpha.
\end{align}
This factor is inserted in \eqref{s2pathintegral}to compensate for the over-counting of gauge equivalent orbits connected by \eqref{lin diff}.


\subsubsection*{Change of variables}

As in the case of massless vectors, we decompose $h_{\mu\nu}$ as 
\begin{align}\label{spin 2 CoV}
	h_{\mu\nu}=h_{\mu\nu}^\text{TT} +\frac{1}{\sqrt{2}} (\nabla_{\mu} \xi_{\nu}+\nabla_{\nu} \xi_{\mu})+\frac{g_{\mu\nu}}{\sqrt{d+1}}\tilde{h}
\end{align} 
where $h_{\mu\nu}^\text{TT}$ is the transverse-traceless part of $h_{\mu\nu}$ satisfying $\nabla^\lambda h_{\lambda \mu}=0=h\indices{^\lambda_\lambda}$, $\xi_{\mu}$ is the pure gauge part of $h_{\mu\nu}$, and $\tilde{h}$ is the trace of $h_{\mu\nu}$. For \eqref{spin 2 CoV} to be unique, we require $\xi_{\nu}$ to be orthogonal to all Killing vectors (KVs) on $S^{d+1}$
\begin{align}\label{xi con}
	(\xi,\xi^\text{KV})=0,\qquad \nabla_{\mu} \xi^\text{KV}_{\nu}+\nabla_{\nu} \xi^\text{KV}_{\mu} = 0
\end{align}
and $\tilde{h}$ to be orthogonal to divergence of the rest of all conformal Killing vectors (CKVs)
\begin{align}\label{tilde h con}
	(\tilde{h},\nabla \cdot\xi^\text{CKV})=0,\qquad \nabla_{\mu} \xi^\text{CKV}_{\nu}+\nabla_{\nu} \xi^\text{CKV}_{\mu} = \frac{1}{2(d+1)}g_{\mu\nu}\nabla^{\lambda}\xi^\text{CKV}_{\lambda}.
\end{align}
The path integral measure then becomes
\begin{align}
	\mathcal{D}h = J \, \mathcal{D}h^\text{TT}\mathcal{D}'\xi \mathcal{D}'\tilde{h}
\end{align}
where the Jacobian $J$ will be found below. The primes indicate that we exclude the integrations over the $(1,1)$ and $(1,0)$ modes excluded due to conditions \eqref{xi con} and \eqref{tilde h con}.

\subsection{Transverse tensor and trace mode determinants}

\subsubsection*{Action for $h^{TT}_{\mu\nu}$}

Due to the gauge invariance \eqref{lin diff}, we have
\begin{align}
	S[h]=S[h^\text{TT}+\tilde{h}]=S[h^\text{TT}]+S[\tilde{h}].
\end{align}
$S[h^\text{TT}]$ can be easily obtained as
\begin{align}
	S[h^\text{TT}]=\frac{1}{2\mathrm{g}^2}\int_{S^{d+1}}h^\text{TT}_{\mu\nu}(-\nabla_{(2)}^2+2)h_\text{TT}^{\mu\nu}.
\end{align}
where $-\nabla_{(2)}^2$ is the spin-2 STT Laplacian. The integration over $h^\text{TT}$ thus gives 
\begin{align}\label{spin2TTaction}
	Z^\text{TT}_h = \int \mathcal{D}h^\text{TT} \, e^{-S[h^\text{TT}]}= \det (-\nabla_{(2)}^2+2)^{-1/2}.
\end{align}

\subsubsection*{Action for $\tilde{h}$ and the conformal factor problem}

Similarly, after a bit more work, the quadratic action for $\tilde{h}$ can be obtained as
\begin{align}\label{spin2tildehaction}
	S[\tilde{h}]=&-\frac{d(d-1)}{2(d+1)\mathrm{g}^2}\int_{S^{d+1}} \tilde{h}(-\nabla_{(0)}^2-(d+1))\tilde{h}\nonumber\\
	=&-\frac{d(d-1)}{2(d+1)\mathrm{g}^2}\sum_{n\neq 1} (n(n+d)-(d+1))c_{n,0}^2
\end{align}
where in the second line we have inserted the mode expansion 
\begin{align}
	\tilde{h}=\sum_{n\neq 1}c_{n,0} f_n,\qquad   ( f_n, f_m)=\delta_{n,m}.
\end{align}
Here the sum runs over the spectrum of the scalar Laplacian except the $(1,0)$ modes, which corresponds to the CKVs. Notice that \eqref{spin2tildehaction} has a wrong overall sign for all positive modes of the operator $-\nabla_{(0)}^2-(d+1)$. This is the well-known conformal factor problem \cite{Gibbons:1978ac} in Euclidean gravity method. We follow the standard prescription: we replace $c_{n,0}\to ic_{n,0}$, \footnote{The sign in front of the $i$ is a matter of convention.} for all $n\geq 2$, which leads to the change in the path integral measure
\begin{align}
	\mathcal{D}'\tilde{h}=\prod_{n\neq 1}\frac{dc_{n,0}}{\sqrt{2\pi}\mathrm{g}}\to\bigg( \prod_{n=2}^\infty i\bigg)\prod_{n\neq 1}\frac{dc_{n,0}}{\sqrt{2\pi}\mathrm{g}}=i^{-d-3}\left(\prod_{n=0}^\infty i\right) \prod_{n\neq 1}\frac{dc_{n,0}}{\sqrt{2\pi}\mathrm{g}}.
\end{align}
The factor in the last step runs through the spectrum of $-\nabla_{(0)}^2$ and is thus a local infinite constant that can be absorbed into bare couplings. Doing this the path integral becomes
\begin{align}
	Z_{\tilde{h}}=&\int  \mathcal{D}'\tilde{h}\, e^{S[\tilde{h}]}= i^{-d-3} Z^+_{\tilde{h}}Z^-_{\tilde{h}} \nonumber\\
	Z^+_{\tilde{h}} =& \int \mathcal{D}^+\tilde{h}\, e^{-\frac{d(d-1)}{2(d+1)\mathrm{g}^2}\int_{S^{d+1}} \tilde{h}(-\nabla_{(0)}^2-(d+1))\tilde{h}}\nonumber\\
	Z^-_{\tilde{h}} =& \int \mathcal{D}^-\tilde{h}\, e^{\frac{d(d-1)}{2(d+1)\mathrm{g}^2}\int_{S^{d+1}} \tilde{h}(-\nabla_{(0)}^2-(d+1))\tilde{h}}
\end{align}
where $\pm$ indicate the contribution from positive and negative modes respectively. The overall phase factor $ i^{-d-3}$ was first obtained by Polchinski \cite{Polchinski:1988ua}. Later we will see the generalization of this phase factor for all massless higher spin fields.

\subsection{Jacobian}\label{spin 2 Jacobian}

Again, we find the Jacobian $J$ by requiring consistency with the normalization condition
\begin{align}
	1=\int\mathcal{D}h \, e^{-\frac{1}{2\mathrm{g}^2}(h,h)}.
\end{align}
Since $h^\text{TT}$ is transverse and traceless, we have
\begin{align}
	(h,h)=(h^\text{TT},h^\text{TT})+(\sqrt{2}\nabla \xi +\frac{g \tilde{h}}{\sqrt{d+1}},\sqrt{2}\nabla \xi +\frac{g \tilde{h}}{\sqrt{d+1}}).
\end{align}
To proceed we separate $\xi_\mu=\xi'_\mu+\xi^{\text{CKV}}_\mu$, where $\xi^{\text{CKV}}_\mu$ is a linear combination of the CKVs and $\xi'_\mu$ is the part of $\xi_\mu$ that is orthogonal to the CKVs, that is $(\xi',\xi^{\text{CKV}})=0$. Note that while $g\tilde{h}$ is orthogonal to $\xi^{\text{CKV}}_\mu$ because of \eqref{tilde h con}, $g\tilde{h}$ and $\nabla \xi'$ are not orthogonal to each other. To remove the off-diagonal terms, we shift
\begin{align}
	\tilde{h}' = \tilde{h}+\sqrt{\frac{2}{d+1}}\nabla^\lambda \xi'_\lambda.
\end{align}
Since it is just a shift, the Jacobian is trivial. It is then easy to compute
\begin{align}
	(\sqrt{2}\nabla \xi +\frac{g \tilde{h}}{\sqrt{d+1}},\sqrt{2}\nabla \xi +\frac{g \tilde{h}}{\sqrt{d+1}})=(\tilde{h}',\tilde{h}')+\frac{1}{2}(K \xi' ,K \xi' )+2(\nabla \xi^\text{CKV},\nabla \xi^\text{CKV})
\end{align}
where we have defined the differential operator
\begin{align}
	(K\xi)_{\mu\nu}\equiv\nabla_\mu \xi_\nu+\nabla_\nu \xi_\mu-\frac{2}{d+1}g_{\mu\nu}\nabla^\lambda \xi_\lambda.
\end{align}
Now the integrations over $h^\text{TT}$ and $\tilde{h}'$ become trivial. To proceed, we first simplify
\begin{align}
	(K \xi' ,K \xi' ) = 2 \int_{S^{d+1}} \bigg[ {\xi'}^\nu \Big( -\nabla^2 -d\Big) \xi'_\nu -{\xi'}^\nu \Big(\frac{d-1}{d+1} \nabla_\nu \nabla^\lambda \xi'_\lambda\Big)\bigg].
\end{align}
Then we decompose $\xi'$ into its transverse and longitudinal parts: $\xi'_\nu =\xi_\nu^T +\nabla_\nu \sigma$. Once again this change of variables leads to a Jacobian factor which is easily found as before. With this decomposition we can further simplify
\begin{align}
	\frac{1}{2}(K \xi' ,K \xi' ) =  S[\xi^T]+\frac{2d}{d+1} S[\sigma],
\end{align}
where 
\begin{align}
	S[\xi^T]=\int_{S^{d+1}} \xi^T_{\nu}(-\nabla^2_{(1)}-d)\xi_T^{\nu},\qquad S[\sigma]=\int_{S^{d+1}} \sigma(-\nabla_{(0)}^2)(-\nabla_{(0)}^2-(d+1))\sigma.
\end{align}
We therefore arrive at
\begin{align}
	\begin{split}\label{spin 2 Jac}
		J=&\frac{W^+_\sigma}{Y^\text{T}_\xi Y^+_\sigma}\frac{1}{Y^\text{CKV}_\xi} \\
		Y^\text{T}_\xi=&\int\mathcal{D}' \xi^T \, e^{-\frac{1}{2\mathrm{g}^2} (\xi^T,(-\nabla^2_{(1)}-d)\xi^T)} \\
		Y^+_\sigma =&\int\mathcal{D}^+\sigma\, e^{-\frac{1}{2\mathrm{g}^2}\frac{2d}{d+1}( \sigma,(-\nabla_{(0)}^2)(-\nabla_{(0)}^2-(d+1))\sigma)}\\
		W^+_\sigma=&\int\mathcal{D}^+\sigma \,e^{-\frac{1}{2\mathrm{g}^2}( \sigma, (-\nabla_{(0)}^2)\sigma)}\\
		Y^\text{CKV}_\xi=&\int\mathcal{D} \xi^\text{CKV}\, e^{-\frac{1}{\mathrm{g}^2}(\nabla \xi^\text{CKV},\nabla \xi^\text{CKV})}
	\end{split}
\end{align}
Here $W^+_\sigma$ is the Jacobian corresponding to the change of variables $\{\xi'_\nu\} \to \{\xi_\nu^T +\nabla_\nu \sigma\}$. The +'s denote the positive modes for the operator $(-\nabla_{(0)}^2-(d+1))$.  \footnote{The zero modes of the operator $(-\nabla_{(0)}^2-(d+1))$ are excluded because $\sigma$ satisfies $(\sigma,f_0)=0=(\sigma,\nabla \xi^\text{CKV})$.}

\subsection{Residual group volume}

As in the massless vector case, we have a factor 
\begin{align}
	\frac{\int \mathcal{D}'\xi}{\text{Vol($\mathcal{G}$)}}
\end{align}
in the path integral. We recall from \eqref{vol g local vec} that Vol($\mathcal{G}$) is a path integral over a local vector field. This does not cancel completely with the integration over $\xi_\mu$, and we are left with a factor (restoring the label $a$ for degenerate modes with same quantum number $(1,1)$)
\begin{align}
	\frac{\int \mathcal{D}'\xi}{\text{Vol($\mathcal{G}$)}}=\frac{1}{\text{Vol}(G)_\text{PI}},\quad \text{Vol}(G)_\text{PI} \equiv \int \prod_{a=1}^{\frac{(d+1)(d+2)}{2}} \frac{d\alpha^{(a)}_{1,1}}{\sqrt{2\pi}\mathrm{g}}. 
\end{align}
where $\alpha^{(a)}_{1,1}$ is the expansion coefficient in the expansion 
\begin{align}
	\alpha_\mu = \sum_{n=1}^\infty \alpha_{n,1} f_{n,\mu}+ \sum_{n=1}^\infty \alpha_{n,0} \hat{T}_{n, \mu}^{(0)}.
\end{align}
These $(1,1)$ modes are diffeomorphisms that leave the background $S^{d+1}$ metric invariant, so they in fact correspond to the Killing vectors of $S^{d+1}$. $G$ is therefore the isometry group $SO(d+2)$ of $S^{d+1}$. As in the massless vector case, we want to relate $\text{Vol}(G)_\text{PI}$ to a canonical volume, following the argument in section \ref{s 1 vol}. 

\paragraph{Local gauge algebra}

Recall that the original Einstein-Hilbert action is invariant under non-linear diffeomorphisms generated by any vector field $\alpha=\frac{1}{\sqrt{2}}\alpha^\mu \partial_\mu$, which reads 
\begin{align}
	\delta_\alpha h_{\mu\nu}=&\delta^{(0)}_\alpha h_{\mu\nu}+\delta^{(1)}_\alpha h_{\mu\nu}+O(h^2)\nonumber\\
	\delta^{(0)}_\alpha h_{\mu\nu}=&\frac{1}{\sqrt{2}}(\nabla_{\mu} \alpha_{\nu}+\nabla_{\nu} \alpha_{\mu})\nonumber\\
	\delta^{(1)}_\alpha h_{\mu\nu}=&\frac{1}{\sqrt{2}}(\alpha^\rho\nabla_\rho h_{\mu\nu}+\nabla_\mu \alpha^\rho h_{\rho\nu}+\nabla_\nu \alpha^\rho h_{\mu\rho}),
\end{align}
where the superscript $(n)$ again denotes the power in fields. This generates the algebra
\begin{align}\label{diff algebra}
	[\delta_\alpha,\delta_{\alpha'}]=\delta_{[[\alpha,\alpha']]}.
\end{align}
In this case, the bracket is proportional to the usual Lie derivative\footnote{If we had worked with canonical normalization, obtained by replacing $h_{\mu\nu}\to \mathrm{g} h_{\mu\nu}$, the bracket will read instead $[[\alpha,\alpha']] =-\frac{\mathrm{g}}{\sqrt{2}}[\alpha,\alpha']_L=-\sqrt{16\pi G_N} [\alpha,\alpha']_L$. This relation can be viewed as a \textit{definition} of the Newton constant $G_N$ in any gauge theory with a massless spin 2 field.}
\begin{align}\label{Einstein bracket}
	[[\alpha,\alpha']] =-\frac{1}{\sqrt{2}}[\alpha,\alpha']_L, \qquad [\alpha,\alpha']_L= (\alpha^\mu\partial_\mu \alpha'^\nu-\alpha'^\mu\partial_\mu \alpha^\nu)\partial_\nu.
\end{align}

\paragraph{Isometry algebra from the local gauge algebra}

The background ($S^{d+1}$) preserving gauge transformations or isometries generated by the Killing vectors satisfying
\begin{align}
	\delta^{(0)}_{\bar{\alpha}}=0
\end{align}
and form a subalgebra of the local gauge algebra, which inherits a bracket from the latter
\begin{align}\label{so(d+2) bracket}
	[[\bar{\alpha},\bar{\alpha}']]=-\frac{1}{\sqrt{2}}[\bar{\alpha},\bar{\alpha}']_L.
\end{align}
To define the canonical volume, we again first find a set of generators $M_{IJ}$ that satisfy the standard $so(d+2)$ commutation relation under the bracket \eqref{so(d+2) bracket}:
\begin{align}
	[[M_{IJ},M_{KL}]]=\eta_{JK}M_{IL}-\eta_{JL}M_{IK}+\eta_{IL}M_{JK}-\eta_{IK}M_{JL}.
\end{align}
One such basis is $M_{IJ}=-\sqrt{2}(X_I\partial_{X^J}-X_J\partial_{X^I})$ where $X^I X_I=1,X^I\in \mathbb{R}^{d+2},I=1\cdots d+2$ are the coordinates of on $S^{d+1}$ represented in the ambient space. Its norm in the invariant bilinear form induced by the path integral is (it suffices to consider only one of the generators)
\begin{align}
	\bra{M_{12}}\ket{M_{12}}_\text{PI}=\frac{1}{2\pi \mathrm{g}^2} \int_{S^{d+1}}(M_{12})^{IJ} (M_{12})_{IJ} =\frac{2}{2\pi \mathrm{g}^2} \int_{S^{d+1}} (X_1^2+X_2^2)=\frac{2}{2\pi \mathrm{g}^2} \frac{2}{d+2}\text{Vol}(S^{d+1}).
\end{align}
Since the canonical bilinear form is defined such that $\bra{M_{12}}\ket{M_{12}}_\text{c}=1$, the path integral metric on $G$ is related to the canonical metric as
\begin{align}
	ds_\text{PI}^2= \frac{2}{2\pi \mathrm{g}^2} \frac{2}{d+2}\text{Vol}(S^{d+1}) ds_\text{can}^2=\frac{1}{8\pi G_N} \frac{\text{Vol}(S^{d-1})}{d(d+2)} ds_\text{can}^2
\end{align}
where we have used $\text{Vol}(S^{d+1})=\frac{2\pi}{d}\text{Vol}(S^{d-1})$ and substituted $\mathrm{g}=\sqrt{32\pi G_N}$ in the last step. Therefore
\begin{align}
	\text{Vol}(G)_\text{PI} =\bigg(\frac{1}{8\pi G_N} \frac{\text{Vol}(S^{d-1})}{d(d+2)} \bigg)^{\frac{(d+1)(d+2)}{4}} \text{Vol}(G)_\text{can} .
\end{align}
The canonical volume $\text{Vol}(G)_\text{can}=\text{Vol}(SO(d+2))_\text{can}$ is well-known:\footnote{This follows from the fact that $SO(n+1)/SO(n)=S^{n}$, which implies that $\text{Vol}(SO(n+1))=\text{Vol}(SO(n))\text{Vol}(S^n)$}
\begin{align}
	\text{Vol}(SO(d+2))_c = \prod_{n=1}^{d+1}\text{Vol}(S^n) =  \prod_{n=1}^{d+1} \frac{2\pi^{\frac{n+1}{2}}}{\Gamma(\frac{n+1}{2})}
\end{align}

\subsection{Final result}

So far we have
\begin{align}\label{spin 2 inter}
	Z_\text{PI}=\frac{i^{-d-3}}{\text{Vol}(G)_\text{PI}} \bigg(\frac{Z^\text{TT}_h}{Y^\text{T}_\xi}\bigg)\bigg(\frac{Z^+_{\tilde{h}} W^+_\sigma}{ Y^+_\sigma}\bigg)\frac{Z^-_{\tilde{h}}}{Y^\text{CKV}_\xi} .
\end{align}
Note that the factor 
\begin{align}
	\frac{Z^\text{TT}_h}{Y^\text{T}_\xi} = \frac{\det'(-\nabla_{(1)}^2-d)^{1/2}}{\det(-\nabla_{(2)}^2+2)^{1/2}} 
\end{align}
is the usual ratio of determinants. Next, the factors in the second bracket in \eqref{spin 2 inter} cancel up to an infinite product
\begin{align}
	\frac{Z^+_{\tilde{h}} W^+_\sigma}{ Y^+_\sigma} =& \int \mathcal{D}^+\tilde{h} \,e^{-\frac{d-1}{4\mathrm{g}^2}\int_{S^{d+1}} \tilde{h}^2} =\prod_{n=2}^\infty \Big( \frac{d-1}{2}\Big)^{-\frac{D^{d+2}_{n,0}}{2}}=\Big( \frac{d-1}{2}\Big)^{\frac{d+3}{2}}\prod_{n=0}^\infty \Big( \frac{d-1}{2}\Big)^{-\frac{D^{d+2}_{n,0}}{2}},
\end{align}
where in the last line we have complete the product so that it runs through the spectrum of the scalar Laplacian. The infinite product can then be absorbed into bare couplings. Finally, the factors in the last bracket in \eqref{spin 2 inter} can be explicitly evaluated to be
\begin{align}
	Z^-_{\tilde{h}}=\bigg(\frac{1}{d(d-1)}\bigg)^{1/2},\quad Y^\text{CKV}_\xi = 2^{-\frac{d+2}{2}}.
\end{align}
Putting everything together, we conclude
\begin{align}
	Z_\text{PI}=&Z_G Z_\text{Char}, \nonumber\\
	Z_G=&i^{-d-3}\frac{\gamma^{\frac{(d+1)(d+2)}{2}}}{ \text{Vol}(SO(d+2))_c },\quad  \gamma=\sqrt{\frac{8\pi G_N}{\text{Vol}(S^{d-1})}}\nonumber\\
	Z_\text{Char}=&\bigg(d(d+2) \bigg)^{\frac{(d+1)(d+2)}{4}}  \frac{(d-1)^{\frac{d+2}{2}}}{(2d)^{1/2}}\frac{\det'(-\nabla_{(1)}^2-d)^{1/2}}{\det(-\nabla_{(2)}^2+2)^{1/2}}.
\end{align}
As a check, we note that except for the inclusion of the phase factor $i^{-d-3}$, for $d=3$ we agree exactly with the 1-loop part of (5.43) in \cite{Volkov:2000ih}.\footnote{Note that our expression agrees with the first line of (5.43) in \cite{Volkov:2000ih}, while the authors made an error in evaluating the determinants, so their second line is incorrect, as already noted in \cite{Anninos:2020hfj}.} In this case one finds \cite{Anninos:2020hfj}
\begin{align} 
	\log Z_{\rm PI}^{S^4} = &\log\left( \frac{i^{-6}}{\frac{2}{3}(2\pi)^6}\left(\frac{8\pi G_N}{4\pi \ell^2}\right)^5 \right) +\frac{8}{3} \frac{1}{\epsilon^4}   \ell^4 -\frac{32}{3} \frac{1}{\epsilon^2}   \ell^2  -\frac{571}{45} \log\left(\frac{2 e^{-\gamma}}{\epsilon} L\right) \nonumber\\
	&- \frac{571}{45}  \log\left(\frac{\ell}{L}\right) 
	+\frac{715}{48} -\log 2 - \frac{47}{3}  \zeta'(-1) +\frac{2}{3}  \zeta'(-3)  
\end{align}
where the sphere radius $\ell$ has been restored. $\epsilon$ is the UV regulator in heat kernel regularization and $L$ is a renormalization scale in the minimal subtraction scheme \cite{Anninos:2020hfj}.


\section{Massless higher spin}\label{Massless PI}

Now we are ready for the 1-loop path integrals for higher spin (HS) theories on $S^{d+1}$. One of the motivations of considering these theories lies in their relevance in holographic proposals \cite{Anninos:2011ui,Anninos:2012ft,Anninos:2013rza,Anninos:2017eib} in de Sitter space. Although the equations of motion for these theories have been constructed \cite{Vasiliev:1990en,Vasiliev:2003ev,Bekaert:2005vh}, the full actions from which these are derived remain elusive.\footnote{See also \cite{Boulanger:2015ova,Sleight:2017pcz} for arguments against the existence for consistent interacting HS theories.} However, since the interactions are at least cubic, their 1-loop partition functions (around the trivial saddle) decouple into a product of free partition functions. 
\begin{align}\label{Vasiliev PI}
	Z^\text{HS}_\text{PI}=\prod_{s} Z^{(s,m^2=0)}_\text{PI},
\end{align}
where $Z^{(s,m^2=0)}_\text{PI}$ is the 1-loop path integral for a massless spin-$s$ field to be described below. The precise range over $s$ in the product depends on the specific higher spin theory we are interested in. In AdS, the determinant expressions for $Z^{(s,m^2=0)}_\text{PI}$ are obtained in \cite{Gaberdiel:2010ar} and \cite{Gupta:2012he}, which are subsequently used in 1-loop tests of HS/CFT dualities \cite{Giombi:2013fka,Giombi:2014iua,Giombi:2016pvg,Gunaydin:2016amv}. In the following, we perform a careful computation for $Z^{(s,m^2=0)}_\text{PI}$ on $S^{d+1}$, whose early stage has some overlap with \cite{Gaberdiel:2010ar}. In fact, the following can be viewed as a derivation for the AdS case as well, except that the latter does not contain the subtleties of phases and group volume that appear on $S^{d+1}$.\footnote{This is because the modes that cause these subtleties are non-normalizable in AdS and are excluded from the beginning.}

\subsection{Operator formalism}

It is much simpler to carry out the entire computation in terms of generating functions, which significantly simplifies tensor manipulations. Here we adopt the convention of \cite{Sleight:2017cax} but on $S^{d+1}$. In this formalism, the tensor structure of a totally symmetric spin-$s$ field $\phi_{\mu_1 \cdots \mu_s}$ in $S^{d+1}$ is encoded in a constant auxiliary ($d+1$)-dimensional vector $u^\mu$:
\begin{align}
	\phi_{(s)}(x)=\phi_{\mu_1 \cdots \mu_s}(x) \to \phi_s(x,u)\equiv \frac{1}{s!} \phi_{\mu_1 \cdots \mu_s}(x) u^{\mu_1} \cdots u^{\mu_s}.
\end{align}
In the following we will suppress the position argument $x$, and interchangeably refer to a rank-$s$ tensor with $\phi_{(s)}$ or its generating function $\phi_s(u)$. Since the original covariant derivative $\nabla_\mu$ acts on both $\phi_{\mu_1 \cdots \mu_s}$ and $u^\mu$, we modify the covariant derivative as
\begin{align}
	\nabla_\mu \to \nabla_\mu +\omega\indices{_\mu ^a ^b}u_a \frac{\partial}{\partial u^b},
\end{align}
where $u^a =e\indices{_\mu^a}u^\mu$ with vielbein $e\indices{_\mu^a}(x)$ and $\omega\indices{_\mu ^a ^b}$ is the spin connection. With this modification the actions of covariant derivatives on $u^\mu$ offset each other, and we can work as if no derivative is acting on $u^\mu$. In the following we will only work in the contracted variables $u^\mu=e\indices{^\mu_a}u^a$ and the associated derivative $\partial_{u^\mu}=e\indices{_\mu^a}\partial_{u^a}$. As a consequence of vielbein postulate we have
\begin{align}
	[\nabla_\mu,u^\nu]=0=[\nabla_\mu,\partial_{u^\nu}].
\end{align}

In this formalism all tensor manipulations are translated to an operator calculus. For instance, tensor contraction:
\begin{align}
	\phi_{\mu_1 \cdots \mu_s} \chi^{\mu_1 \cdots \mu_s} =s! \phi_s(\partial_u)\chi_s(u).
\end{align}
In particular, the inner product \eqref{inner} is represented as
\begin{align}
	(  \phi_s, \chi_s ) =s! \int_{S^{d+1}} \phi_s(\partial_u)\chi_s(u).
\end{align}
List of operations:
\begin{align}\label{op list}
	\text{divergence: }& \nabla \cdot \partial_u, & \text{sym. gradient: }&  u\cdot \nabla, & \text{Laplacian: }& \nabla^2,\nonumber \\
	\text{sym. metric: }& u^2, & \text{trace: }&  \partial_u^2, & \text{spin: }& u\cdot \partial_u.
\end{align}
One of the biggest advantages of this formalism is that we can work algebraically with these operators without explicitly referring to the tensor. For example, to define the de Donder operator, we can either state explicitly its action on a spin-$s$ field $\phi_{(s)}$
\begin{align}
	\hat{D}\phi_{(s)}=\hat{D}\phi_{\mu_1 \cdots \mu_s} =  \nabla^\lambda \phi_{\mu_1 \cdots \mu_{s-1}\lambda} -\frac{1}{2} \nabla_{(\mu_1}\phi\indices{_{\mu_2 \cdots \mu_{s-1})\lambda}^\lambda}
\end{align}
or simply in terms of its generating function
\begin{align}
	\hat{D}(\nabla,u,\partial_u)=& \nabla \cdot \partial_u -\frac{1}{2}(u\cdot \nabla) (\partial_u^2).
\end{align}
In the following we will use these two kinds of notations interchangeably.

On $S^{d+1}$, the operators \eqref{op list} satisfy the following operator algebra
\begin{align}
	[\nabla_\mu, \nabla_\nu] =& u_\mu\partial_{u^\nu}-u_\nu \partial_{u^\mu} \label{algebra1}\\
	[\nabla^2 , u\cdot\nabla] =& u\cdot \nabla (2u\cdot\partial_u +d)-2u^2 \nabla\cdot \partial_u \label{algebra2}\\
	[\nabla\cdot \partial_u , \nabla^2] =& (2u\cdot\partial_u +d)  \nabla\cdot \partial_u-2 u\cdot \nabla \partial_u^2\label{algebra3}\\
	[\nabla\cdot \partial_u , u\cdot\nabla]=& \nabla^2 + u\cdot \partial_u (u\cdot \partial_u+d-1)-u^2 \partial_u^2\label{algebra4} \\
	[\nabla\cdot \partial_u ,u^2] =& 2u\cdot \nabla\label{algebra5}\\
	[ \partial_u^2 ,u\cdot\nabla] =& 2 \nabla\cdot \partial_u\label{algebra6}\\
	[\partial_u^2,u^2]=& 2(d+1+2u\cdot \partial_u)\label{algebra7}
\end{align}
where we have denoted $\partial_u^2 \equiv \partial_u \cdot \partial_u, u^2 \equiv u\cdot u$.


\subsection{Fronsdal action on $S^{d+1}$}

The 1-loop partition function for a free bosonic spin-$s$ massless gauge field on $S^{d+1}$ is
\begin{align}\label{eq:Epathintegral}
	Z^{(s)}_\text{PI}=\frac{1}{\text{Vol($\mathcal{G}_{s}$)}}\int \mathcal{D}\phi_{(s)}\,e^{-S[\phi_{(s)}]}
\end{align}
where the quadratic Fronsdal action \cite{Fronsdal:1978rb} in the operator language is given by
\begin{align}\label{eq:opaction}
	S[\phi_{(s)}] = \frac{s!}{2\mathrm{g}_s^2}\int_{S^{d+1}} \phi_s (\partial_u)\Big(1-\frac{1}{4}u^2 \partial_u^2 \Big)\hat{\mathcal{F}}_s (\nabla,u,\partial_u) \phi_s (u)
\end{align}
with $\hat{\mathcal{F}}_s(\nabla,u,\partial_u) $ is the Fronsdal operator
\begin{align}
	\hat{\mathcal{F}}_s(\nabla,u,\partial_u) =&-\nabla^2+M_s^2-u^2 \partial_u^2 +u\cdot \nabla \hat{D}(\nabla,u,\partial_u)\\
	\hat{D}(\nabla,u,\partial_u)=& \nabla \cdot \partial_u -\frac{1}{2}(u\cdot \nabla) (\partial_u^2),
\end{align}
where 
\begin{align}\label{mass para}
	M_s^2=s-(s-2)(s+d-2)
\end{align}
and $\hat{D} $ is the de Donder operator. An $s$-dependent factor $\mathrm{g}_s^2$ is inserted as an overall factor. Canonical normalization corresponds to setting $\mathrm{g}_s=1$. We will choose a particular value for $\mathrm{g}_s$ when we discuss the issue of group volume. (\ref{eq:opaction}) is invariant under the gauge transformations
\begin{align}\label{eq:opgauge}
	\phi_s (u)\mapsto \phi_s (u) + \frac{1}{\sqrt{s}}u\cdot\nabla\Lambda_{s-1}(u).
\end{align}
In this off-shell formalism $\phi_s (u)$ satisfies a double-tracelessness condition (trivial for $s\leq 3$)
\begin{align}\label{eq:opdouble-tracelessness}
	(\partial_u^2)^2 \phi_s (u)=0 ,
\end{align}
which implies that the gauge parameter $\Lambda_{(s-1)}$ must be traceless (imposed even for $s=3$)
\begin{align}
	\partial_u^2 \Lambda_{s-1}(u)=0 .
\end{align}
The division by the gauge group volume $\text{Vol($\mathcal{G}_{s}$)}$ in \eqref{eq:Epathintegral} compensates for the overcounting of gauge equivalent configurations connected by \eqref{eq:opgauge}.

\subsubsection*{Change of variables}

To proceed, we change field variables
\begin{align}\label{opsdisplit}
	\phi_s (u) = \phi^\text{TT}_s (u)+ \frac{1}{\sqrt{s}}u\cdot \nabla \xi_{s-1}(u) +\frac{1}{\sqrt{2s(s-1)(d+2s-3)}}u^2 \chi_{s-2}(u).
\end{align}
Here $\phi_{(s)}^\text{TT}$ is the transverse traceless piece of $\phi_{(s)}$ for which
\begin{align}
	\nabla\cdot\partial_u \phi^\text{TT}_s (u)=&0= \partial_u^2 \phi^\text{TT}_s (u).
\end{align}
Next, $\xi_{(s-1)}$ is the symmetric traceless spin-($s-1$) gauge parameters which are required to be orthogonal to all spin-($s-1$) Killing tensors $\epsilon_{(s-1)}^{KT}$ (which generate trivial gauge transformations) so that it is uniquely fixed:
\begin{align}\label{xi condition}
	\partial_u^2 \xi_{s-1}(u) =&0,\quad (\xi_{(s-1)}, \epsilon_{(s-1)}^{KT})=0
\end{align}
Finally, $\chi_{(s-2)}$ is the spin-($s-2$) piece which carries all the trace information of $\phi_{(s)}$. The double-tracelessness condition (\ref{eq:opdouble-tracelessness}) implies that $\chi_{(s-2)}$ is traceless:
\begin{align}\label{chi traceless}
	\partial_u^2 \chi_{s-2}(u) =&0.
\end{align}
Note that if $\epsilon^{CKT}_{(s-1)}$ is a conformal Killing tensor (CKT) satisfying 
\begin{align}\label{CKeq}
	\hat{K}_{s} (\nabla,u,\partial_u)\epsilon^{CKT}_{ s-1}(u)\equiv u\cdot \nabla \epsilon^{CKT}_{ s-1}(u)-\frac{u^2}{d+2s-3}(\nabla \cdot \partial_u )\epsilon^{CKT}_{ s-1}(u)=0,
\end{align}
then any new set of variables related by the transformation
\begin{align}
	\xi_{s-1}(u)&\to\xi_{s-1}(u)+\epsilon^{CKT}_{s-1}(u)\\
	\chi_{s-2}(u)&\to\chi_{s-2}(u)-\frac{u^2}{d+2s-3}(\nabla \cdot \partial_u )\epsilon^{CKT}_{ s-1}(u)
\end{align}
will result in the same $\phi_{(s)}$. To uniquely fix $\chi_{(s-2)}$, we thus impose
\begin{align}\label{eq:gauge}
	( \chi_{(s-2)},(\nabla \cdot\epsilon^{CKT})_{(s-2)} )=0
\end{align}
for all the spin-($s-1$) CKTs $\epsilon^{CKT}_{(s-1)}$. The path integral measure then becomes
\begin{align}\label{sdimeasure}
	\mathcal{D}\phi_s = J_{(s)}  \mathcal{D}\phi^\text{TT}_{(s)}\mathcal{D}'\xi_{(s-1)} \mathcal{D}'\chi_{(s-2)}
\end{align}
where the Jacobian $J_{(s)}$ will be found below. The primes indicate that we exclude the $(s-1,m)$ ($0\leq m\leq s-1$) modes excluded due to conditions \eqref{xi condition} and \eqref{eq:gauge}.

\subsection{Quadratic actions for $\phi^\text{TT}_{(s)}$ and $\chi_{(s-2)}$}

\subsubsection*{Action for $S[\phi_{(s)}^\text{TT}] $}

The quadratic action for $\phi^\text{TT}_{(s)}$ is
\begin{align}
	S[\phi_{(s)}^\text{TT}] =\frac{s!}{2\mathrm{g}_s^2} \int_{S^{d+1}} \phi^\text{TT}_s (\partial_u) ( -\nabla_{(s)}^2+M_s^2 )\phi^\text{TT}_s (u)=\frac{1}{2\mathrm{g}_s^2}( \phi_{(s)}^\text{TT}, ( -\nabla_{(s)}^2+M_s^2) \phi_{(s)}^\text{TT} ).
\end{align}
which leads to the path integral
\begin{align}
	Z^{(s)}_{\phi^\text{TT}}=\int \mathcal{D}\phi_{(s)}^\text{TT}\, e^{-\frac{1}{2\mathrm{g}_s^2} ( \phi_{(s)}^\text{TT}, ( -\nabla_{(s)}^2+M_s^2) \phi_{(s)}^\text{TT} )}=\det(-\nabla_{(s)}^2+M_s^2)^{-1/2}
\end{align}

\subsubsection*{Action for $S[\chi_{(s-2)}]$ and the HS conformal factor problem}

From (\ref{eq:opaction}) we have
\begin{align}\label{eq:op pre tilde phi action}
	S[\chi_{(s-2)}] =& \frac{(s-2)!}{8\mathrm{g}_s^2} \int_{S^{d+1}}  \chi_{s-2}(\partial_u)(\partial_u^2)\Big(1-\frac{1}{4}u^2 \partial_u^2 \Big)\hat{\mathcal{F}}_s (\nabla,u,\partial_u) u^2 \chi_{s-2}(u)\nonumber\\
	=&-\frac{(s-2)!(d+2s-5)}{8(d+2s-3)\mathrm{g}_s^2}\int_{S^{d+1}}\chi_{s-2}(\partial_u)(\partial_u^2) \hat{\mathcal{F}}_s (\nabla,u,\partial_u) u^2 \chi_{s-2}(u)
\end{align}
where we have used \eqref{algebra7} and the tracelessness of $ \chi_{(s-2)}$ \eqref{chi traceless}. Using the operator algebras, one easily finds that
\begin{align}
	&\hat{\mathcal{F}}_s (\nabla,u,\partial_u) u^2 \nonumber\\
	=&u^2 \Big(- \nabla^2-s(-1+d+s)+2\Big)+ u^2 (u\cdot\nabla)( \nabla\cdot \partial_u)-(d+2s-5)(u\cdot\nabla)^2+\cdots
\end{align}
where and henceforth $\cdots$ denotes terms that will not contribute because of the tracelessness condition \eqref{chi traceless}: $\partial_u^2 \chi_{s-2}(u) =0$ or $\chi_{s-2}(\partial_u) u^2 =0$. Then we have
\begin{align}
	(\partial_u^2)\hat{\mathcal{F}}_s (\nabla,u,\partial_u) u^2 = &4(d+2s-4) \Big(- \nabla^2-(s-1)(s+d-2)-1\Big)  \nonumber\\
	&-2(d+2s-7) (u\cdot\nabla)( \nabla\cdot \partial_u)+\cdots.
\end{align}
Defining the differential operator
\begin{align}\label{Qop}
	\hat{\mathcal{Q}}(\nabla,u,\partial_u)\equiv &2\frac{d+2s-4}{d+2s-3}\Big(-\nabla^2 -(s-1)(s+d-2)-1\Big)-\frac{d+2s-7}{d+2s-3}(u\cdot\nabla) (\nabla\cdot\partial_u),
\end{align}
the quadratic action for $\chi_{(s-2)}$ is simply
\begin{align}
	S[\chi_{(s-2)}] =&-\frac{d+2s-5}{4\mathrm{g}_s^2} (\chi_{(s-2)}, \hat{\mathcal{Q}} \chi_{(s-2)} ).
\end{align}
To proceed, we expand $\chi_{(s-2)}$ (see appendix \ref{STSH} for the properties of the induced symmetric traceless spherical harmonics)
\begin{align}
	\chi_{(s-2)} = \sum_{m=0}^{s-2}  A_{s-2,m} \hat{T}_{s-2, (s-2)}^{(m)}+ \sum_{m=0}^{s-2} \sum_{n=s}^\infty A_{n,m} \hat{T}_{n, (s-2)}^{(m)},
\end{align}
where the modes $(n,m)=(s-1,m), 0\leq m\leq s-2$ are excluded because of the condition \eqref{eq:gauge}. It is easy to verify that $\hat{\mathcal{Q}}$ is negative for the modes in the first sum and positive in the second. This is the HS generalization of the conformal factor problem. To make the integrals converge, we replace $A_{n,m}\to iA_{n,m}$ for $0\leq m\leq s-2, s\leq n<\infty$, leading to the change in the path integral measure
\begin{align}
	\mathcal{D}'\chi_{(s-2)}  = \prod_{m=0}^{s-2}  \frac{dA_{s-2,m}}{\sqrt{2\pi}\mathrm{g}_s}\prod_{m=0}^{s-2} \prod_{n=s}^\infty \frac{dA_{n,m}}{\sqrt{2\pi}\mathrm{g}_s}\to \bigg(\prod_{m=0}^{s-2} \prod_{n=s}^\infty i^{D^{d+2}_{n,m}}\bigg) \prod_{m=0}^{s-2}  \frac{dA_{s-2,m}}{\sqrt{2\pi}\mathrm{g}_s}\prod_{m=0}^{s-2} \prod_{n=s}^\infty \frac{dA_{n,m}}{\sqrt{2\pi}\mathrm{g}_s}.
\end{align}
We complete the product so that it runs through the spectrum for the unconstrained spin-($s-2$)  Laplacian, i.e. 
\begin{align}
	\prod_{m=0}^{s-2} \prod_{n=s}^\infty i^{D^{d+2}_{n,m}}=i^{-N^\text{CKT}_{s-2}-N^\text{CKT}_{s-1} +N^\text{KT}_{s-1}}\prod_{m=0}^{s-2} \prod_{n=s-2}^\infty i^{D^{d+2}_{m,n}}
\end{align}
where $N^\text{CKT}_{s}=\sum_{m=0}^{s} D^{d+2}_{s,m}$ and $N^\text{KT}_{s}=D^{d+2}_{s,s}$ are the number of spin-$s$ CKTs and spin-$s$ KTs respectively. The local infinite product can then be absorbed into bare couplings. The remaining phase factor is the HS generalization of the Polchinski's phase. We can then write the path integral over $\chi_{(s-2)}$ as
\begin{align}\label{phis-2}
	Z^{(s)}_\chi=&i^{-N^\text{CKT}_{s-2}-N^\text{CKT}_{s-1} +N^\text{KT}_{s-1}} Z^{(s)}_{\chi^+}Z^{(s)}_{\chi^-}\nonumber\\
	Z^{(s)}_{\chi^+}=&\int  \mathcal{D}^+\chi_{(s-2)}\,e^{-\frac{d+2s-5}{4\mathrm{g}_s^2}( \chi_{(s-2)},\hat{\mathcal{Q}}\chi_{(s-2)})}\nonumber\\
	Z^{(s)}_{\chi^-}=&\int\mathcal{D}^-\chi_{(s-2)}\, e^{\frac{d+2s-5}{4\mathrm{g}_s^2}( \chi_{(s-2)},\hat{\mathcal{Q}}\chi_{(s-2)})} 
\end{align}
where the superscripts $\pm$ denotes integrations over the positive (negative) modes of $\hat{\mathcal{Q}}$.

\subsection{Jacobian}

Again, we find the Jacobian in \eqref{sdimeasure} by the normalization condition
\begin{align}\label{snorm}
	\int \mathcal{D} {\phi}_{(s)} \, e^{-\frac{1}{2\mathrm{g}_s^2}( {\phi}_{(s)},{\phi}_{(s)})}=1.
\end{align}
We plug in (\ref{sdimeasure}) and \eqref{opsdisplit} to find $J_{(s)}$. Notice that $\phi^\text{TT}_{(s)}$ is orthogonal to $g\chi_{(s-2)}$ and $\nabla \xi_{(s-1)}$ with respect to the inner product $( \cdot , \cdot )$; on the other hand, when $\xi$'s are orthogonal to the spin-($s-1$) CKTs (denoted as $\xi'$), $g\chi_{(s-2)}$ and $\nabla \xi'_{(s-1)}$ are not orthogonal, and we remove the off-diagonal terms by shifting
\begin{align}\label{furtherCoV}
	\chi'_{s-2}(u) =\chi_{s-2}(u) +\sqrt{\frac{s(s-1)}{2(d+2s-3)}}(\nabla \cdot \partial_u ) \xi'_{s-1}(u).
\end{align}
The Jacobian corresponding to this shift is trivial. We then have
\begin{align}
	( \phi_{(s)},\phi_{(s)} )= ( \phi^\text{TT}_{(s)},\phi^\text{TT}_{(s)} )+( \chi'_{(s-2)},\chi'_{(s-2)})+\frac{1}{s}( \hat{K}_{s}\xi'_{(s-1)} , \hat{K}_{s} \xi'_{(s-1)} )+\frac{1}{s}( \nabla\xi^{CKT}_{(s-1)} , \nabla\xi^{CKT}_{(s-1)} ).
\end{align}
where $\hat{K}_{s}$ is the operator appearing in \eqref{CKeq}. It is useful to note that acting on any symmetric traceless tensor $\epsilon_{(s-1)}$,
\begin{align}\label{K iden}
	\partial_u^2 \hat{K}_{s}(\nabla,u,\partial_u)\epsilon_{s-1}(u)=0=\hat{K}_{s}(\nabla,\partial_u,u)\epsilon_{s-1}(\partial_u)u^2.
\end{align}
The path integrals over $\phi^\text{TT}_{(s)}$ and $\chi'_{(s-2)}$ are trivial, and therefore $J_{(s)}$ can be expressed as
\begin{align}
	J_{(s)}^{-1} =& Y^{(s)}_{\xi'} Y^{(s)}_{\xi^\text{CKT}} \\
	Y^{(s)}_{\xi'}\equiv &\int \mathcal{D}\xi'_{(s-1)}\, e^{-\frac{1}{2s\mathrm{g}_s^2}( K\xi'_{(s-1)} , K\xi'_{(s-1)} )}\\
	Y^{(s)}_{\xi^\text{CKT}} \equiv&\int \mathcal{D}\xi^{CKT}_{(s-1)}\, e^{-\frac{1}{2s\mathrm{g}_s^2}( \nabla\xi^{CKT}_{(s-1)} , \nabla\xi^{CKT}_{(s-1)} )}.
\end{align}

\subsubsection*{Expressing $Y_{\xi'}$ in terms of functional determinants}

To proceed, we use the operator algebra and \eqref{K iden} and simplify
\begin{align}
	\frac{1}{s}( \hat{K}_{s}\xi'_{(s-1)} , \hat{K}_{s} \xi'_{(s-1)} ) =& ( \xi'_{(s-1)}, \Big( -\nabla_{(s-1)}^2-(s-1)(s+d-2) \Big)\xi'_{(s-1)} ) \nonumber\\
	&+\frac{d+2s-5}{d+2s-3}( \xi'_{(s-1)},-\nabla \nabla\boldsymbol{\cdot}\xi'_{(s-1)}).
\end{align}
We then perform the change of variables
\begin{align}\label{xiCoV}
	\xi'_{(s-1)} = {\xi'}^{\text{TT}}_{(s-1)}+\hat{K}_{s-1}\sigma_{(s-2)},
\end{align}
where ${\xi'}^{\text{TT}}_{(s-1)}$ is the transverse traceless part of ${\xi'}_{(s-1)}$, $\sigma_{(s-2)}$ is a spin-($s-2$) symmetric traceless field and the differential operator $\hat{K}_{s-1}(\nabla,u,\partial_u)$ is defined in \eqref{CKeq}. We require $\sigma_{(s-2)}$ to be orthogonal to the kernel of $\hat{K}_{s-1}$, i.e. the spin-($s-2$) CKTs. Also, ${\xi'}^{\text{TT}}_{(s-1)}$ and $\sigma_{(s-2)}$ are automatically orthogonal to the spin-($s-1$) CKTs.

Plugging in these, we have two decoupled pieces
\begin{align}\label{gh inter}
	\frac{1}{s}( \hat{K}_{s}\xi'_{(s-1)} , \hat{K}_{s} \xi'_{(s-1)} ) =& S[ {\xi'}_{(s-1)}^\text{TT}]+S[\sigma_{(s-2)}].
\end{align}
Here the first term is the ghost action
\begin{align}
	S[ {\xi'}_{(s-1)}^\text{TT}]=& ( {\xi'}_{(s-1)}^\text{TT}, \Big( -\nabla_{(s-1)}^2+m_{s-1,s}^2+M_{s-1}^2\Big){\xi'}_{(s-1)}^\text{TT} ) \label{gh STT}
\end{align}
with $M_{s-1}^2$ as defined in \eqref{mass para} and we have defined 
\begin{align}\label{PM mass}
	m_{s,t}^2=(s-1-t)(d+s+t-3),
\end{align}
which is exactly the mass for a partially massless field with spin-$s$ and depth $t$ for $0\leq t\leq s-1$. The second term in \eqref{gh inter} is the action of a spin-$(s-2)$ field
\begin{align}
	S[\sigma_{(s-2)}]=& ( \hat{K}_{s-1}\sigma_{(s-2)}, \hat{\mathcal{P}}\hat{K}_{s-1}\sigma_{(s-2)}) \label{sigma}\\
	\hat{\mathcal{P}}(\nabla,u,\partial_u)=&-\nabla_{(s-1)}^2-(s-1)(s+d-2)-\frac{d+2s-5}{d+2s-3}(u\cdot\nabla)( \nabla\cdot\partial_u)
\end{align}
To proceed, we commute $\hat{\mathcal{P}}$ and $\hat{K}_{s-1}$. This requires the relation
\begin{align}
	\nabla_{(s-1)}^2 \hat{K}_{s-1}(\nabla,u,\partial_u) - \hat{K}_{s-1}(\nabla,u,\partial_u)\nabla_{(s-2)}^2 = (d+2s-4)u\cdot \nabla+\cdots
\end{align}
and the commutator
\begin{align}
	&[(u\cdot \nabla )(\nabla\cdot \partial_u),\hat{K}_{s-1}(\nabla,u,\partial_u)] \nonumber\\
	=& [(u\cdot \nabla )(\nabla\cdot \partial_u),u\cdot\nabla]-\frac{1}{d+2s-5}[(u\cdot \nabla )(\nabla\cdot \partial_u), u^2 \nabla \cdot \partial_u]
\end{align}
which can be computed using
\begin{align}
	[(u\cdot \nabla )(\nabla\cdot \partial_u),u\cdot\nabla] =& (u\cdot \nabla ) \Big(\nabla^2 +(s-2)(s+d-3) \Big)+\cdots \\
	[(u\cdot \nabla )(\nabla\cdot \partial_u), u^2 \nabla \cdot \partial_u]=&2(u\cdot \nabla )^2 (\nabla\cdot \partial_u)+\cdots 
\end{align}
where and henceforth $\cdots$ denotes terms that will not contribute to \eqref{sigma} because of the tracelessness condition \eqref{K iden} of the operator $\hat{K}_{s-1}$. We have also used the fact that $u\cdot \partial_u \sigma_{s-2}(u)=(s-2)\sigma_{s-2}(u)$. To briefly summarize,
\begin{align}
	& \hat{\mathcal{P}}(\nabla,u,\partial_u)\hat{K}_{s-1}(\nabla,u,\partial_u) \nonumber\\
	=& \hat{K}_{s-1} (\nabla,u,\partial_u)\hat{\mathcal{P}}(\nabla,u,\partial_u)-(d+2s-4)u\cdot \nabla\nonumber\\
	&+\frac{d+2s-5}{d+2s-3}(u\cdot \nabla )\bigg[  \Big(-\nabla^2 -(s-2)(s+d-3) \Big)+\frac{2}{d+2s-5}(u\cdot \nabla ) (\nabla\cdot \partial_u)\bigg]+\cdots .
\end{align}
Now, observe that because of \eqref{K iden}, $u\cdot \nabla$ can be replaced by the operator $\hat{K}_{s-1}$
\begin{align}
	u\cdot \nabla = \hat{K}_{s-1}(\nabla,u,\partial_u) +\cdots 
\end{align}
up to trace terms that do not contribute to \eqref{sigma}. Therefore we have
\begin{align}
	\hat{\mathcal{P}}(\nabla,u,\partial_u)\hat{K}_{s-1}(\nabla,u,\partial_u) = \hat{K}_{s-1}(\nabla,u,\partial_u)\hat{\mathcal{W}}(\nabla,u,\partial_u)+\cdots,
\end{align}
with
\begin{align}
	\hat{\mathcal{W}}(\nabla,u,\partial_u)=&\hat{\mathcal{P}}(\nabla,u,\partial_u)-(d+2s-4)\nonumber\\
	&+\frac{d+2s-5}{d+2s-3}\bigg[  \Big(-\nabla^2 -(s-2)(s+d-3) \Big)+\frac{2}{d+2s-5}(u\cdot \nabla ) (\nabla\cdot \partial_u)\bigg]
\end{align}
Amazingly, one can show that this operator is exactly equal to $\hat{\mathcal{Q}}$ defined in \eqref{Qop}, that is $\hat{\mathcal{W}}(\nabla,u,\partial_u)=\hat{\mathcal{Q}}(\nabla,u,\partial_u)$. So we have found
\begin{align}
	S[\sigma_{(s-2)}]=( \hat{K}_{s-1}\sigma_{(s-2)}, \hat{K}_{s-1}\hat{\mathcal{Q}}\sigma_{(s-2)})= (\sigma_{(s-2)},  \hat{K}_{s-1}^\dagger\hat{K}_{s-1}\hat{\mathcal{Q}}\sigma_{(s-2)}).
\end{align}
To conclude, we have 
\begin{align}
	Y^{(s)}_{\xi'}=&\frac{Y^{(s)}_{\xi^\text{TT}}Y^{(s)}_{\sigma^+}}{ W^{(s)}_{\sigma^+}} \label{Yxi}\\
	Y^{(s)}_{\xi^\text{TT}}\equiv&\int\mathcal{D} {\xi'}^{\text{TT}}_{(s-1)}\, e^{-\frac{1}{2\mathrm{g}_s^2} ( {\xi'}_{(s-1)}^\text{TT}, \Big( -\nabla_{(s-1)}^2+m_{s-1,s}^2+M_{s-1}^2 \Big){\xi'}_{(s-1)}^\text{TT} )}\\
	Y^{(s)}_{\sigma^+}\equiv &\int \mathcal{D}^+\sigma_{(s-2)}\, e^{-\frac{1}{2\mathrm{g}_s^2}( \sigma_{(s-2)}, K^\dagger K \hat{\mathcal{Q}}\sigma_{(s-2)})}\\
	W^{(s)}_{\sigma^+}=&\int \mathcal{D}^+\sigma_{(s-2)}\, e^{-\frac{1}{2\mathrm{g}_s^2}( \sigma_{(s-2)}, K^\dagger K\sigma_{(s-2)})}
\end{align}
Here the superscript $+$ emphasizes the fact that we are integrating over modes orthogonal to the spin-($s-1$) and spin-($s-2$) CKTs. In particular, this is the part of spectrum that coincides with the ``+'' integral in \eqref{phis-2}. Here $(W^{(s)}_{\sigma^+})^{-1}$ is the Jacobian associated with the change of variables \eqref{xiCoV}.

\subsection{Residual group volume}

Recall that after the change of variables \eqref{opsdisplit}, the integration over the pure gauge modes $\xi$ decoupled from the $\phi^\text{TT}_{(s)}$ and $\chi_{(s-2)}$ path integrals, and we are left with a factor (we have restored the label $a$ for degenerate modes with same quantum number $(s-1,s-1)$)
\begin{align}
	\frac{\int\mathcal{D}' \xi_{(s-1)}}{\text{Vol($\mathcal{G}_{s}$)}}=\frac{1}{\text{Vol($G_{s}$)}},\quad \text{Vol($G_{s}$)} =\int \prod_{a=1}^{N_{s-1}^\text{KT}}\frac{d\alpha^{(a)}_{s-1,s-1}}{\sqrt{2\pi}\mathrm{g}_s}.
\end{align}
due to the integration over the spin-($s-1$) Killing tensor modes. This leads to a product in the original path integral \eqref{Vasiliev PI}:
\begin{align}\label{HS PI vol}
	\text{Vol($G$)}_\text{PI}=\prod_{s} \text{Vol($G_{s}$)}.
\end{align}
HS symmetries typically form infinite dimensional groups. Therefore there is an issue of making sense of \eqref{HS PI vol}, which we are not going to attempt in this paper. 

\paragraph{HS invariant bilinear form}

Instead, we are going to do a more modest task. As in the warm-up examples, the volume $\text{Vol($G$)}_\text{PI}$ is defined with a particular metric, namely
\begin{align}\label{hs PI metric}
	ds_\text{PI}^2=\frac{1}{2\pi } \sum_{s} \frac{1}{\mathrm{g}_s^2} d\alpha_{s-1,s-1}^2.
\end{align}
Again we want to express this in terms of a canonical metric with respect to which we define a canonical volume $\text{Vol($G$)}_\text{can}$. There are however complications compared to the massless spin-1 and spin-2 cases:
\begin{enumerate}
	\item As opposed to the case for Yang-Mills or Einstein gravity, we do not know the full nonlinear actions for Vasiliev theories that give rise to the interacting equations of motion and the full nonlinear gauge transformations in the metric-like formalism. This implies that we do not know the full local HS gauge algebra. Fortunately, the global part of the algebra does not require this knowledge, but only the lowest order ones, which only requires the information of the cubic couplings. 
	
	\item Another complication is that since HS symmetries mix different spins, the HS invariant bilinear form depends on the relative normalizations of fields withe different spins in the action. Once this is fixed, the bilinear form is uniquely determined up to an overall normalization.
\end{enumerate}
All of these have been worked out in the case of a negative cosmological constant \cite{Joung:2013nma}. To go to the case of a positive cosmological constant is a simple matter of analytic continuation. In appendix \ref{cubic}, we translate the relevant results from \cite{Joung:2013nma} to the case of $S^{d+1}$. The final result is that upon choosing
\begin{align}
	\mathrm{g}_s^2=s!,
\end{align}
the HS invariant bilinear form is determined to be
\begin{align}\label{HS can form}
	\bra{\bar{\alpha}_1}\ket{\bar{\alpha}_2}_\text{can} =\frac{8\pi G_N}{\text{Vol}(S^{d-1})}\sum_{s} (d+2s-2)(d+2s-4) \bra{\bar{\alpha}_{1,(s-1)}}\ket{\bar{\alpha}_{2,(s-1)}}_\text{PI}
\end{align}
where the overall normalization is again fixed by requiring the canonical spin-2 generators to be unit-normalized with respect to \eqref{HS can form}. This implies that the group volume \eqref{HS PI vol} is related to the canonical volume as
\begin{align}\label{HS can vol}
	\text{Vol($G$)}_\text{PI}=\text{Vol($G$)}_\text{can}\prod_s \left( \frac{\text{Vol}(S^{d-1})}{8\pi G_N}\frac{1}{(d+2s-2)(d+2s-4)}\right)^{\frac{N^\text{KT}_{s-1}}{2}}.
\end{align}

\subsection{Final result}

So far we have
\begin{align}
	Z^\text{HS}_\text{PI}=& \frac{i^{-P}}{\text{Vol($G$)}_\text{PI}}\prod_{s}
	\Bigg(\frac{Z^{(s)}_{\phi^\text{TT}}}{Y^{(s)}_{\xi^\text{TT}}}\Bigg)\Bigg( \frac{Z^{(s)}_{\chi^+}W^{(s)}_{\sigma^+}}{Y^{(s)}_{\sigma^+}}\Bigg)\Bigg( \frac{Z^{(s)}_{\chi^-}}{ Y^{(s)}_{\xi^\text{CKT}} }\Bigg)
\end{align}
where $P=\sum_s (N^\text{CKT}_{s-2}+N^\text{CKT}_{s-1} -N^\text{KT}_{s-1})$. In the infinite product, the first factor is the usual ratio of determinants of physical and ghost operators
\begin{align}
	\frac{Z^{(s)}_{\phi^\text{TT}}}{Y^{(s)}_{\xi^\text{TT}}}=\frac{\det'(-\nabla_{(s-1)}^2+m_{s-1,s}^2+M_{s-1}^2)^{1/2}}{ \det(-\nabla_{(s)}^2+M_s^2)^{1/2}} .
\end{align}
In the second factor, $Z^+_{\chi}, W^+_\sigma, Y^+_{\sigma}$ run over the exact same spectrum and cancel almost completely up to an infinite constant 
\begin{align}
	\frac{Z^+_{\chi}W^+_\sigma}{Y^+_{\sigma}}=&\int  \mathcal{D}\chi^+_{(s-2)}\,e^{-\frac{d+2s-5}{4\mathrm{g}_s}( \chi^+_{(s-2)},\chi^+_{(s-2)})}\nonumber\\
	=&\frac{\int  \mathcal{D}\chi_{(s-2)}\,e^{-\frac{d+2s-5}{4\mathrm{g}_s}( \chi_{(s-2)},\chi_{(s-2)})}}{\int  \mathcal{D}\chi^0_{(s-2)}\mathcal{D}\chi^-_{(s-2)}\,e^{-\frac{d+2s-5}{4\mathrm{g}_s}( \chi_{(s-2)},\chi_{(s-2)})}}
\end{align}
where in the denominator $\chi^0_{(s-2)}$ denotes the modes excluded due to \eqref{eq:gauge}. The infinite constant in the numerator is a path integral over the entire spectrum of an unconstrained spin-$(s-2)$ symmetric traceless field and therefore can be absorbed into bare couplings. To proceed, we plug in explicit mode expansions 
\begin{align}
	\chi^0_{(s-2)}=&\sum_{m=0}^{s-1}  A_{s-1,m} \hat{T}_{s-1, (s-2)}^{(m)},\quad \chi^-_{(s-2)}=&\sum_{m=0}^{s-2}  A_{s-2,m} \hat{T}_{s-2, (s-2)}^{(m)},\quad \xi^{CKT}_{(s-1)}  =&\sum_{m=0}^{s-2}  A_{s-1,m}\hat{T}_{s-1, (s-1)}^{(m)},
\end{align}
which lead to 
\begin{align}
	\frac{Z^{(s)}_{\chi^+}W^{(s)}_{\sigma^+}}{Y^{(s)}_{\sigma^+}}=& \prod_{m=0}^{s-2} \prod_{n=s-2}^{s-1} \bigg[\frac{ 2}{d+2s-5}\bigg]^{D^{d+2}_{n,m}/2}\\
	Z^{(s)}_{\chi^-}=&\prod_{m=0}^{s-2} \Big[\frac{2}{(d+2s-5)m_{s+1,m}^2}\Big]^{\frac{D^{d+2}_{s-2,m}}{2}}\\
	Y^{(s)}_{\xi^\text{CKT}}=&\prod_{m=0}^{s-2} \bigg[ \frac{2 m^2_{s,m}}{d+2 s-5}\bigg]^{-\frac{D^{d+2}_{s-1,m}}{2}}.
\end{align}
We therefore have
\begin{align}\label{finite fac}
	\Bigg( \frac{Z^{(s)}_{\chi^+}W^{(s)}_{\sigma^+}}{Y^{(s)}_{\sigma^+}}\Bigg)\Bigg( \frac{Z^{(s)}_{\chi^-}}{ Y^{(s)}_{\xi^\text{CKT}} }\Bigg)=&\prod_{m=0}^{s-2}(m_{s+1,m}^2)^{-\frac{D^{d+2}_{s-2,m}}{2}} \prod_{m=0}^{s-2} ( m^2_{s,m})^{\frac{D^{d+2}_{s-1,m}}{2}}.
\end{align}
Together with the determinant factor, this can be further written as
\begin{align}
	\Bigg(\frac{Z^{(s)}_{\phi^\text{TT}}}{Y^{(s)}_{\xi^\text{TT}}}\Bigg)\Bigg( \frac{Z^{(s)}_{\chi^+}W^{(s)}_{\sigma^+}}{Y^{(s)}_{\sigma^+}}\Bigg)\Bigg( \frac{Z^{(s)}_{\chi^-}}{ Y^{(s)}_{\xi^\text{CKT}} }\Bigg)
	=\frac{\det\nolimits'_{-1}|-\nabla_{(s-1)}^2-\lambda_{s-1,s-1}|^{1/2}}{ \det\nolimits'_{-1}|-\nabla_{(s)}^2-\lambda_{s-2,s}|^{1/2}}.
\end{align}
Here the subscript $-1$ means that we extend the eigenvalue product from $n=s$ to $n=-1$. The primes denote omission of the zero modes from the determinants. In the numerator we omitted the $n=s-1$ mode while in the denominator we omitted the $n=s-2$ mode.\footnote{Originally $\lambda_{n,s}$ and $D^{d+2}_{n,s}$ were defined only for $n\geq s$, which are now extended to all $n\in \mathbb{Z}$.} To obtain this expression we used the relation
\begin{align}\label{eigen PM}
	\lambda_{t-1,s}+M_s^2 =-m_{s,t}^2
\end{align}
and the fact that $D^{d+2}_{s-1,t}=-D^{d+2}_{t-1,s}$ (implying $D^{d+2}_{s-1,s}=0$). This extension of the eigenvalue product from $n=s$ to $n=-1$ is exactly the prescription described in \cite{Anninos:2020hfj}. Putting everything together, we finally obtain the expression
\begin{align}\label{HS final}
	Z^\text{HS}_\text{PI}=&Z_\text{G}Z_\text{Char}\nonumber\\
	Z_\text{G}=&i^{-P} \frac{\gamma^{\text{dim}\,G}}{\text{Vol($G$)}_\text{can}},\qquad Z_\text{Char}=\prod_s Z^{(s)}_\text{Char}\nonumber\\
	Z^{(s)}_\text{Char}=& \left( \frac{(d+2s-2)(d+2s-4)}{M^4}\right)^{\frac{N^\text{KT}_{s-1}}{2}}\frac{\det\nolimits'_{-1}\left|\frac{-\nabla_{(s-1)}^2-\lambda_{s-1,s-1}}{M^2}\right|^{1/2}}{ \det\nolimits'_{-1}\left|\frac{-\nabla_{(s)}^2-\lambda_{s-2,s}}{M^2}\right|^{1/2}},
\end{align}
with 
\begin{gather}
	P=\sum_s (N^\text{CKT}_{s-2}+N^\text{CKT}_{s-1} -N^\text{KT}_{s-1}), \qquad \gamma=\sqrt{\frac{8\pi G_N}{\text{Vol}(S^{d-1})}},\qquad \text{dim}\,G =\sum_s N^\text{KT}_{s-1}
\end{gather}
Note that we have restored the dimensionful parameter $M$. As noted in \cite{Anninos:2020hfj}, the factor $(d+2s-2)(d+2s-4)$ gets nicely canceled after evaluating the character integrals for the determinants.


\section{Massive fields}\label{Massive PI}

Now let us turn to fields with generic masses. In this case we do not have a group volume factor, and thus no coupling dependence. We will work with canonical normalizations. 

\subsection{Massive scalars and vectors}

\paragraph{Massive scalars}

The path integral for a scalar $\phi$ with mass $m^2>0$ is simply
\begin{align}
	Z^{(s=0,m^2)}_\text{PI}=\int \mathcal{D}\phi\, e^{-\frac{1}{2}\int_{S^{d+1}}\phi (-\nabla^2+m^2)\phi}=\det(-\nabla^2+m^2)^{-1/2}
\end{align}

\paragraph{Massive vectors}

Massive vectors are described by the Proca action
\begin{align}\label{Proca action}
	S[A]=&\int_{S^{d+1}} \Big(\frac{1}{4}F_{\mu\nu}F^{\mu\nu} +\frac{m^2}{2} A_\mu A^\mu\Big).
\end{align}
Similar to the massless case, to proceed we make a change of variables \eqref{spin 1 CoV} with Jacobian \eqref{vector Jac}, so that the action becomes
\begin{gather}
	S[A]=S[A^T]+S[\chi]\nonumber\\
	S[A^T]=\frac{1}{2}(A^T, (-\nabla_{(1)}^2+m^2+d)A^T),\quad S[\chi] =\frac{m^2}{2}(\chi,(-\nabla_{(0)}^2)\chi).
\end{gather}
For $m^2>0$ that corresponds to unitary de Sitter representations, the result is
\begin{align}
	Z^{(s=1,m^2)}_\text{PI}=\det(-\nabla_{(1)}^2+m^2+d)^{-1/2} (m^2)^{1/2}=\det\nolimits_{-1}(-\nabla_{(1)}^2+m^2+d)^{-1/2}.
\end{align}
The presence of the factor $(m^2)^{1/2}$ originates from the fact that the $(0,0)$ mode is excluded from the integration over the longitudinal mode. In the last equality we again note that the multiplication of the factor $(m^2)^{1/2}$ is equivalent to extending the product to $n=-1$.

\subsection{Massive spin 2 and beyond}

\subsubsection{Massive $s=2$}

The action for a free massive spin-2 field on $S^{d+1}$ is (see for example \cite{Higuchi:1986py})
\begin{align}\label{massivespin2action}
	S[h] = &\frac{1}{2}\int_{S^{d+1}}h^{\mu\nu}\bigg[ (-\nabla^2+2)h_{\mu\nu}+2\nabla_{(\mu}\nabla^\lambda h_{\nu) \lambda}+g_{\mu\nu}(\nabla^2 h\indices{_\lambda^\lambda}-2\nabla^{\sigma}\nabla^\lambda h_{\sigma \lambda})+(d-2)g_{\mu\nu}h\indices{_\lambda^\lambda}\nonumber\\
	&+m^2(h_{\mu\nu}-g_{\mu\nu}h\indices{_\lambda^\lambda})\bigg].
\end{align}
If we put $m=0$ we recover the action \eqref{eq:spin2action} (with $\mathrm{g}=1$) for linearized gravity. To proceed, we again change the variables \eqref{spin 2 CoV}. It is convenient to further decompose $\xi_\mu$ into its transverse and longitudinal parts: $\xi_\mu=\xi_\mu^T+\nabla_\mu \sigma$, so that the full decomposition for $h_{\mu\nu}$ is
\begin{align}\label{spin 2 full CoV}
	h_{\mu\nu}=&h_{\mu\nu}^\text{TT} +\frac{1}{\sqrt{2}} (\nabla_{\mu} \xi^T_{\nu}+\nabla_{\nu} \xi^T_{\mu}) +\sqrt{2}\nabla_\mu\nabla_\nu \sigma +\frac{g_{\mu\nu}}{\sqrt{d+1}} \tilde{h} .
\end{align}
For this decomposition to be unique, we impose
\begin{align}\label{new xi con}
	(\xi^T,f_{1,(1)})=0\quad , \quad (\tilde{h},f_1)=0\quad  \text{and} \quad (\sigma,f_0)=0.
\end{align}
The first two constraints are equivalent to \eqref{xi con} and \eqref{tilde h con} while the last one ensures $\nabla_\mu \sigma\neq 0$. With a slight modification of the steps in section \ref{spin 2 Jacobian}, the Jacobian for the \eqref{spin 2 full CoV} is obtained as
\begin{align}
	\begin{split}\label{spin 2 full Jac}
		\mathcal{D}h =& J  \mathcal{D}h^\text{TT}\mathcal{D}'\xi^T\mathcal{D}^+\sigma \mathcal{D}'\tilde{h}\\
		J=&\frac{1}{Y^\text{T}_\xi Y^+_\sigma Y^\text{CKV}_\xi} \\
		Y^\text{T}_\xi=&\int\mathcal{D}' \xi^T \,e^{-\frac{1}{2} (\xi^T,(-\nabla^2_{(1)}-d)\xi^T)} \\
		Y^+_\sigma =&\int\mathcal{D}^+\sigma \,e^{-\frac{1}{2}\frac{2d}{d+1}( \sigma,(-\nabla_{(0)}^2)(-\nabla_{(0)}^2-(d+1))\sigma)}\\
		Y^0_\sigma=&\int\mathcal{D}^0 \sigma\, e^{-(\sigma,(-\nabla_{(0)}^2)(-\nabla_{(0)}^2-d) \sigma)}=\int\mathcal{D}^0 \sigma \,e^{-(d+1)(\sigma, \sigma)}.
	\end{split}
\end{align}
Here $\mathcal{D}^+\sigma$ ($\mathcal{D}^0\sigma$) involves integrations over only the positive (zero) modes for the operator $(-\nabla_{(0)}^2-(d+1))$. After substituting \eqref{spin 2 full CoV} the action decouples into 
\begin{align}
	S[h]=S[h^{TT}]+S[\xi^{T}]+S[\sigma,\tilde{h}].
\end{align}
The quadratic actions for $h^\text{TT}$ and $\xi^T$ are simply
\begin{align}\label{mass TT action}
	S[h^\text{TT}]=\frac{1}{2}\int_{S^{d+1}}h^\text{TT}_{\mu\nu}(-\nabla_{(2)}^2+m^2+2)h_\text{TT}^{\mu\nu},
\end{align}
and
\begin{align}\label{mass transverse action}
	S[\xi^T]=\frac{m^2}{2} ( \xi^T , (-\nabla_{(1)}^2 -d)\xi^T)
\end{align}
respectively. Since $\sigma$ and $\tilde{h}$ are not orthogonal, they mix in the action
\begin{align}\label{spin 2 scalar action}
	S[\sigma,\tilde{h}] =& -\frac{(d-1)d}{2(d+1)}( \tilde{h}, (-\nabla_{(0)}^2+(d+1)(\frac{ m^2}{d-1}-1))\tilde{h} ) -\sqrt{\frac{2}{d+1}} d m^2 ( \nabla_{(0)}^2 \sigma , \tilde{h}) \nonumber \\
	&+\frac{m^2}{2} ( \nabla \sigma , (-\nabla_{(0)}^2 -d)\nabla \sigma) -\frac{m^2}{2} ( \nabla_{(0)}^2 \sigma,\nabla_{(0)}^2 \sigma ).
\end{align}
To diagonalize $S[\sigma,\tilde{h}]$, we make a shift (with a trivial Jacobian)\footnote{Because of the constraints \eqref{new xi con}, the $(0,0)$ and $(1,0)$ modes do not mix in \eqref{spin 2 scalar action}}
\begin{align}\label{spin 2 shift}
	\sigma'=\sigma-\frac{1}{\sqrt{2(d+1)}}\tilde{h}
\end{align}
for all scalar modes $f_n$ with $n\geq 2$, so that $S[\sigma,\tilde{h}] =S[\sigma',\tilde{h}] =S[\sigma'] +S[\tilde{h}]$, with 
\begin{align}\label{spin 2 de scalar action}
	S[\sigma'] =-d m^2 (  \sigma',-\nabla_{(0)}^2 \sigma' )\quad \text{and} \quad S[\tilde{h}]=\frac{d(m^2-(d-1))}{2(d+1)}(  \tilde{h},(-\nabla_{(0)}^2-(d+1))  \tilde{h} ).
\end{align}
Notice that $S[\sigma']$ and $S[\tilde{h}]$ vanishes identically when $m^2 =0$ and $m^2 =d-1$ respectively. These are the cases when we have gauge symmetries. The massless case has already been discussed in section \ref{massless spin2}. The case of $m^2 =d-1$ will be considered in section \ref{PM fields}.

Depending on the precise value of $m^2>-2(d+2)$,\footnote{This is the range where the kinetic operator in \eqref{mass TT action} is positive definite. The case $m^2<-2(d+2)$ will be considered when we discuss the shift-symmetric spin-2 fields in section \ref{shift sym}.} some of the modes in \eqref{mass transverse action} and \eqref{spin 2 de scalar action} might acquire an overall negative sign. We Wick rotate the negative modes, absorbing local infinite constants into bare couplings. This will induce a phase factor. Below we give a summary for different cases (n.m. stands for negative modes):
\renewcommand{\arraystretch}{1.5}
\begin{center}
	\begin{tabular}{ |c|c|c|c|c| } 
		\hline
		Range of $m^2$ & n.m. in $S[\xi^T]$ & n.m. in $S[\sigma']$ &n.m. in $S[\tilde{h}]$ & Phase \\
		\hline
		$-2(d+2)<m^2<0$ &$f_{n,\mu}, n\geq 1$ & None & $f_n, n\geq 2$ & $i^{-D^{d+2}_{1,1}-D^{d+2}_{1,0}}=i^{-\frac{(d+3)(d+2)}{2}}$\\ 
		\hline
		$0<m^2<d-1$ & None & $f_n,n\geq 1$ & $f_n,n\geq 2$ & $i^{-2D^{d+2}_{0,0}-D^{d+2}_{1,0}}=i^{-d-4}$\\ 
		\hline
		$m^2>d-1$ & None & $f_n,n\geq 1$ & $f_0$ & $i^{0}=1$\\ 
		\hline
	\end{tabular}
\end{center}
The last case ($m^2>d-1$) is precisely the case when the corresponding de Sitter representations are unitary.\footnote{Principal series for $m^2>(\frac{d}{2})^2$ and complementary series for $d-1<m^2<(\frac{d}{2})^2$ \cite{Basile:2016aen}.} We will focus on this case from now on.

Putting everything together, we have
\begin{align}
	Z^{(s=2,m^2)}_\text{PI}=Z^\text{TT}_h\bigg(\frac{Z^\text{T}_\xi}{Y^\text{T}_\xi}\bigg)\bigg(\frac{Z^+_{\sigma'}Z^+_{\tilde{h}}}{ Y^+_\sigma}\bigg)\bigg(\frac{Z^0_{\sigma'}}{Y^0_\sigma}\bigg)Z^-_{\tilde{h}}
\end{align}
Here $Z^\text{TT}_h,Z^\text{T}_\xi,Z^\pm_{\sigma'},Z^0_{\sigma'},Z^\pm_{\tilde{h}}$ are the path integrals with actions \eqref{mass TT action}, \eqref{mass transverse action} and \eqref{spin 2 de scalar action}. The labels $\pm$ and 0 denote the positive (negative) and zero modes for the scalar operator $-\nabla_{(0)}^2-(d+1)$. Every factor can be easily evaluated:
\begin{align}
	\begin{split}
		Z^\text{TT}_h =& \det (-\nabla_{(2)}^2+m^2+2)^{-1/2} \\
		\frac{Z^\text{T}_\xi}{Y^\text{T}_\xi} =& \int\mathcal{D}' \xi^T\, e^{-\frac{m^2}{2} (\xi^T,\xi^T)}\\
		\frac{Z^+_{\sigma'}Z^+_{\tilde{h}}}{ Y^+_\sigma}=&\int\mathcal{D}^+\sigma' \,e^{-\frac{m^2}{2}  (  \sigma', \sigma' )}\int\mathcal{D}^+\tilde{h}\,e^{-\frac{d(m^2-(d-1))}{2}(  \tilde{h},  \tilde{h} )}\\
		\frac{Z^0_{\sigma'}}{Y^0_\sigma}=&\int\mathcal{D}^0\sigma' \,e^{-\frac{d m^2}{2} (  \sigma', \sigma' )}\\
		Z^-_{\tilde{h}}=&\int\mathcal{D}^-\tilde{h} \,e^{-\frac{d(m^2-(d-1))}{2}(  \tilde{h},  \tilde{h} )}.
	\end{split}
\end{align}
Observe that all factors but $Z^\text{TT}_h$ can be combined in the following way:
\begin{align}
	\bigg(\frac{Z^\text{T}_\xi}{Y^\text{T}_\xi}\bigg)\bigg(\frac{Z^+_{\sigma'}Z^+_{\tilde{h}}}{ Y^+_\sigma}\bigg)\bigg(\frac{Z^0_{\sigma'}}{Y^\text{CKV}_\xi}\bigg)Z^-_{\tilde{h}}=&\frac{\int\mathcal{D} \xi \,e^{-\frac{m^2}{2} (\xi,\xi)}\int\mathcal{D}\tilde{h} \,e^{-\frac{d(m^2-(d-1))}{2}(  \tilde{h},  \tilde{h} )}}{\int\mathcal{D}^0\sigma'\, e^{-\frac{ (m^2-(d-1))}{2} (  \sigma', \sigma' )}\int\mathcal{D}^0 \xi^T \,e^{-\frac{m^2}{2} (\xi^T,\xi^T)}}.
\end{align}
In the numerator, the path integrations are over local unconstrained fields and thus can be absorbed into bare couplings. In the denominator $\mathcal{D}^0 \xi^T$ denotes integration over the modes $f_{1,\mu}$. The integrals in the denominator can be easily evaluated. To conclude, we have
\begin{align}
	Z^{(s=2,m^2)}_\text{PI}=&\det (-\nabla_{(2)}^2+m^2+M_2^2)^{-1/2}(m^2-m_{2,0}^2)^{\frac{D^{d+2}_{1,0}}{2}}(m^2-m_{2,1}^2)^{\frac{D^{d+2}_{1,1}}{2}}\nonumber\\
	=&\det\nolimits_{-1} (-\nabla_{(2)}^2+m^2+M_2^2)^{-1/2}
\end{align}
where we recall that $m_{s,t}^2$ is defined in \eqref{PM mass}.

\subsubsection{Massive arbitrary spin $s\geq 1$}

In principle, one starts with the full manifestly local and covariant action \cite{Zinoviev:2001dt}, which involves a tower of spin $t< s$ Stueckelberg fields, and repeat the derivation above. However, having worked out the cases for $s=1,2$, the pattern is clear. For a free massive spin-$s$ field, its path integral is simply
\begin{align}\label{massive general}
	Z^{(s,m^2)}_\text{PI}=\det\nolimits_{-1} \left(\frac{-\nabla_{(s)}^2+m^2+M_s^2}{M^2}\right)^{-1/2}.
\end{align}
Note that we have restored the dimensionful parameter $M$. Recall that the scaling dimension $\Delta$ is related to the mass $m^2$ as
\begin{align}\label{mass dim}
	m^2 =(\Delta+s-2)(d+s-2-\Delta)
\end{align}
so that
\begin{align}
	\lambda_{n,s}+m^2+M_s^2=(n+\Delta)(d+n-\Delta)=\left(n+\frac{d}{2}\right)^2-\left(\Delta-\frac{d}{2}\right)^2.
\end{align}
The requirement that $\lambda_{n,s}+m^2+M_s^2$ is positive for all $n\geq -1$ is equivalent to the unitary bounds on $\Delta$ \cite{Basile:2016aen}:
\begin{align}
	\Delta=\frac{d}{2}+i \nu,\nu\in\mathbb{R} \quad \text{(Principal series)} \quad \text{or} \quad 1<\Delta <d-1\quad \text{(Complementary series)}
\end{align}
Outside of this bound, a finite number of $\lambda_{n,s}+m^2+M_s^2$ will become negative, which leads to the presence of some power of $i$, as we have seen in the $s=2$ case. Also, as we take $m^2 \to m_{s,t}^2$, \eqref{massive general} becomes ill-defined, signaling a gauge symmetry. The case of $t=s-1$ is the massless case discussed in section \ref{Massless PI}. We will comment on the general $(s,t)$ case in section \ref{PM fields}.


\section{Shift-symmetric fields}\label{shift sym}

In (A)dS space, when massive fields attain certain mass values, they can have shift symmetries  \cite{Bonifacio:2018zex} that generalize the shift symmetry, galileon symmetry, and special galileon symmetry of massless scalars in flat space. In AdS, these theories are unitary; in dS, these theories do not fall into the classifications of dS UIRs \cite{Basile:2016aen}.\footnote{However, there are arguments (see e.g. \cite{Bonifacio:2018zex}) that they can be cured to become unitary.} In the following we study their 1-loop (free) path integrals on $S^{d+1}$, which contain analogous subtleties as the massless case, namely the phases and group volumes. 

\subsection{Shift-symmetric scalars}

Let us start with a free scalar $\phi$ with generic mass $m$, with action 
\begin{align}\label{massive s0 action}
	S[\phi]=\frac{1}{2}\int_{S^{d+1}}\phi (-\nabla^2+m^2)\phi.
\end{align}
When $m^2$ takes values of the negative the eigenvalues of the scalar Laplacian $-\nabla_{(0)}^2$, i.e.
\begin{align}
	m^2=-\lambda_{k,0}=-k(k+d)=m_{0,k+1}^2+M_0^2=m_{k+2,1}^2=-m_{2,k+1}^2\leq 0,\quad k\geq 0,
\end{align}
(recall that $m_{s,t}^2$ is defined in \eqref{PM mass}), the action is invariant under a shift symmetry (of level $k$ in the terminology of \cite{Bonifacio:2018zex})
\begin{align}
	\delta \phi = f_k
\end{align}
where $f_k$ is the $(k,0)$ eigenmodes of $-\nabla^2_{(0)}$ with eigenvalue $\lambda_k$. 

\subsubsection{$k=0$: massless scalars}

The simplest example is $k=0$, i.e. a massless scalar \cite{Allen:1985ux,Allen:1987tz}: 
\begin{align}
	S[\phi]=\frac{1}{2}\int_{S^{d+1}}\phi (-\nabla^2)\phi
\end{align}
which is invariant under a constant shift $\phi \to \phi +c$. The case with $d=3$ is of particularly interest because of its relevance in inflation. The path integral is simply
\begin{align}
	Z^{(s=0,m^2=0)}_\text{PI} =\text{Vol}(G)_\text{PI}\det\nolimits'\left(-\nabla_{(0)}^2\right)^{-1/2}\, .
\end{align}
Here the prime denotes the omission of the constant $(0,0)$ mode from the functional determinant. The group volume factor is the integral over the constant mode
\begin{align}
	\text{Vol}(G)_\text{PI}\equiv \int \frac{dA_{0,0}}{\sqrt{2\pi}}.
\end{align}
Unlike the case of massless gauge fields, the residual group volume is multiplying the determinant instead of being divided. As for massless gauge fields, $\text{Vol}(G)_\text{PI}$ depends on the non-linear completion of the theory. There will be a problem of relating $\text{Vol}(G)_\text{PI}$ to a canonical volume $\text{Vol}(G)_\text{can}$ and the determination of $\text{Vol}(G)_\text{can}$ itself. Also, we expect there will be a dependence on coupling constants of the interacting theory.

An example for which we can make sense of these issues is that of a compact scalar. They are scalars subject to the identification
\begin{align}
	\phi \sim \phi +2\pi R\,, 
\end{align}
that is, they take values on a circle of radius $R$. In this case the integration range for the (0,0) mode is restricted to the fundamental domain $0<A_{0,0}<2\pi R \sqrt{\text{Vol}(S^{d+1})}$ and therefore
\begin{align}
	Z^{\text{compact scalar}}_\text{PI}=\sqrt{2\pi R^2\text{Vol}(S^{d+1})}\det\nolimits'(-\nabla_{(0)}^2)^{-1/2}.
\end{align}
Here the (inverse of) radius $R$ plays the role of the coupling constant.

\subsubsection{$k\geq 1$: tachyonic scalars}

For any $k\geq 1$, the scalar is tachyonic. See for example \cite{Bros:2010wa} and \cite{Epstein:2014jaa} for the study of such tachyonic scalars. The $k=1$ and $k=2$ cases are the dS analogs for the Galileon and special Galileon theories in flat space \cite{Bonifacio:2018zex} respectively. Note that the action is negative for all $(n,0)$ modes with $n<k$, and vanishes for the $(k,0)$ modes. To make sense of the path integral, we again perform Wick rotations for all $(n,0)$ modes with $n<k$, so that
\begin{align}
	\int \mathcal{D}^{<k}\phi \, e^{-S_{<k}[\phi]} \to i^{\sum_{n=0}^{k-1}D^{d+2}_{n,0}} \int \mathcal{D}^{<k}\phi\, e^{S_{<k}[\phi]}=i^{\sum_{n=0}^{k-1}D^{d+2}_{n,0}}\prod_{n=0}^{k-1}|\lambda_{n,0}-\lambda_{k,0}|^{-1/2},
\end{align}
and interpret the integration over the $(k,0)$ modes as a residual group volume
\begin{align}
	\text{Vol}(G_k)_\text{PI}\equiv \int \prod_{a=1}^{D^{d+2}_{k,0}}\frac{dA^{(a)}_{k,0}}{\sqrt{2\pi}}.
\end{align}
The modes with $n>k$ can be integrated as usual. The final result is
\begin{align}
	Z^{(s=0,m_{k+2,1}^2)}_\text{PI}=i^{\sum_{n=0}^{k-1}D^{d+2}_{n,0}}\text{Vol}(G_k)_\text{PI}\det\nolimits'|-\nabla_{(0)}^2-\lambda_{k,0}|^{-1/2}\,.
\end{align}
Note that absolute value is taken in the determinant. The prime denotes the omission of the $(k,0)$ modes from the functional determinant. Same as the $k=0$ case, the residual group volume $\text{Vol}(G_k)_\text{PI}$ is multiplying the determinant instead of being divided. Again, $\text{Vol}(G_k)_\text{PI}$ should depend on the interaction structure of the parent theory. There will be a problem of relating $\text{Vol}(G_k)_\text{PI}$ to a canonical volume $\text{Vol}(G_k)_\text{can}$ and the determination of $\text{Vol}(G_k)_\text{can}$ itself. 

\subsection{Shift-symmetric vectors}

When the mass takes values 
\begin{align}
	m^2=-\lambda_{k+1,1}-d=-(k+2)(k+d)=m_{k+3,0}^2=-m_{1,k+2}^2\leq -2d,\quad k\geq 0,
\end{align}
the Proca action \eqref{Proca action} is invariant under a level-$k$ shift symmetry generated by the $(k+1,1)$ modes
\begin{align}
	\delta A_{\mu} = f_{k+1,\mu}.
\end{align}
Following analogous steps as for the scalars, it is straightforward to work out the path integral
\begin{align}
	Z^{(s=1,m_{k+3,0}^2)}_\text{PI}=i^{\sum_{n=-1}^{k}D^{d+2}_{n,1}}\text{Vol}(G_{k+1,1})_\text{PI}\det'\nolimits_{-1}|-\nabla_{(1)}^2-\lambda_{k+1,1}|^{-1/2} 
\end{align}
where the prime denotes the omission of the $(k+1,1)$ modes and
\begin{align}
	\text{Vol}(G_{k+1,1})_\text{PI}\equiv\int\prod_{a=1}^{D^{d+2}_{k+1,1}} \frac{dA^{(a)}_{k+1,1}}{\sqrt{2\pi}}.
\end{align}
Note that in the phase factor we have used the fact that $D^{d+2}_{-1,1}=-D^{d+2}_{0,0}$ and $D^{d+2}_{0,1}=0$.

\subsection{Shift-symmetric spin $s\geq 2$}

\subsubsection{Shift-symmetric spin 2 fields}

The massive spin-2 action \eqref{massivespin2action} with 
\begin{align}
	m^2=-\lambda_{k+2,2}-2=m_{k+4,1}^2=-m_{2,k+3}^2\leq 2(d+2),\quad k\geq 0,
\end{align}
is invariant under a level-$k$ shift symmetry generated by the $(k+2,2)$ modes
\begin{align}
	\delta h_{\mu\nu} = f_{k+2,\mu\nu}.
\end{align}
It is straightforward to work out the path integral
\begin{align}
	Z^{(s=2,m_{k+4,1}^2)}_\text{PI}=i^{\sum_{n=-1}^{k+1}D^{d+2}_{n,2}}\text{Vol}(G_{k+2,2})_\text{PI}\det'\nolimits_{-1}|-\nabla_{(2)}^2-\lambda_{k+2,2}|^{-1/2} 
\end{align}
where the prime denotes the omission of the $(k+2,2)$ modes and
\begin{align}
	\text{Vol}(G_{k+2,2})_\text{PI}\equiv\int\prod_{a=1}^{D^{d+2}_{k+2,2}} \frac{dA^{(a)}_{k+2,2}}{\sqrt{2\pi}}.
\end{align}
Note that in the phase factor we have used the fact that $D^{d+2}_{-1,2}=-D^{d+2}_{1,0}$ and $D^{d+2}_{0,2}=-D^{d+2}_{1,1}$.

\subsubsection{Shift-symmetric arbitrary spins $s\geq 0$}

Now the pattern is clear. When the mass for a spin-$s$ field $\phi_{(s)}$ ($s\geq 0$) reaches the values
\begin{align}
	m^2=-\lambda_{k+s,s}^2-M_s^2=m_{k+s+2,s-1}^2=-m_{s,s+k+1}^2,\quad k\geq 0,
\end{align}
there will be a level-$k$ shift symmetry generated by the $(k+s,s)$ modes
\begin{align}
	\delta \phi_{(s)} = f_{k+s,(s)}.
\end{align}
The path integral is 
\begin{align}\label{shift general}
	Z^{(s,m_{k+s+2,s-1}^2)}_\text{PI}=i^{\sum_{n=-1}^{k+s-1}D^{d+2}_{n,s}}\text{Vol}(G_{k+s,s})_\text{PI}\det'\nolimits_{-1}\left|\frac{-\nabla_{(s)}^2-\lambda_{k+s,s}}{M^2}\right|^{-1/2} 
\end{align}
where
\begin{align}
	\text{Vol}(G_{k+s,s})_\text{PI}\equiv\int\prod_{a=1}^{D^{d+2}_{k+s,s}} \frac{M}{\sqrt{2\pi}}dA^{(a)}_{k+s,s}.
\end{align}
Note that we have restored the dimensionful parameter $M$. Such a shift-symmetric field can be thought of as the longitudinal mode decoupled from a massive spin-$(k+s+1)$ field as its mass approaches $m_{k+s+1,s}^2$. Note that for $k=0$, it can be thought of as the ghost part of the spin-$(s+1)$ massless path integral. We will see more connections of shift-symmetric fields with general partially massless fields in the next section.


\section{Partially massless fields}\label{PM fields}

In (A)dS space, there exist ``partial massless'' (PM) representations \cite{Higuchi:1986py,Zinoviev:2001dt,Deser:1983tm,DESER1984396,Brink:2000ag,Deser:2001pe,Deser:2001us, Deser:2001wx,Deser:2001xr, Skvortsov:2006at, Hinterbichler:2016fgl}. Except for the massless case, they are not unitary in AdS. In $dS_{d+1}$ with $d\geq 4$, they correspond to the unitary exceptional series representations, while for $d=3$ they correspond to the discrete series representations \cite{Basile:2016aen}. A PM spin-$s$ field of depth $t$ has a gauge symmetry\footnote{We adopt the convention that depth $t$ is equal to the spin of the gauge parameter.}
\begin{align}
	\delta \phi_{(s)}=\nabla^{(s-t)}\xi_{(t)}+\cdots 
\end{align}
where $\cdots$ stand for terms with fewer derivatives \cite{Hinterbichler:2016fgl}. The massless case corresponds to $t=s-1$. In the following we first work out the case of spin-2 depth-0 field. Then we will provide a general prescription for general PM fields.

\subsection{Spin-2 depth-0 field}

The action for a spin-2 depth-0 field is \eqref{massivespin2action} with mass
\begin{align}
	m^2=m_{2,0}^2=d-1,
\end{align}
in which case there is a gauge symmetry 
\begin{align}
	\delta h_{\mu\nu}=\nabla_\mu\nabla_\nu \chi+g_{\mu\nu}\chi.
\end{align}
This can be seen by first substituting \eqref{spin 2 shift} into \eqref{spin 2 full CoV} so that the decomposition becomes
\begin{align}
	h_{\mu\nu}=&h_{\mu\nu}^\text{TT} +\frac{1}{\sqrt{2}} (\nabla_{\mu} \xi^T_{\nu}+\nabla_{\nu} \xi^T_{\mu}) +\sqrt{2}\nabla_\mu\nabla_\nu \sigma' +\frac{1}{\sqrt{d+1}} \left(\nabla_\mu\nabla_\nu\tilde{h}+g_{\mu\nu}\tilde{h}\right)
\end{align}
and noting that $S[\tilde{h}]$ defined in \eqref{spin 2 de scalar action} vanishes identically for $m^2=d-1$. Spin-2 field with such a mass was first considered in \cite{Higuchi:1986py}. This gauge invariance implies that there is an integration
\begin{align}
	\int\mathcal{D}'\tilde{h}
\end{align}
that must be canceled by a gauge group volume factor $\text{Vol}(\mathcal{G})$ divided by hand. To be consistent with locality, this gauge group factor must take the form of a path integral of a local scalar field $\alpha$
\begin{align}
	\text{Vol}(\mathcal{G}) =\int \mathcal{D}\alpha.
\end{align}
Due to mismatch of modes excluded due to \eqref{new xi con}, we have a residual group volume
\begin{align}
	\frac{ \int\mathcal{D}'\tilde{h}}{\text{Vol}(\mathcal{G})}=\frac{1}{\text{Vol}(G_{1,0})_\text{PI}},\quad \text{Vol}(G_{1,0})_\text{PI} \equiv \int\prod_{a=1}^{D^{d+2}_{1,0}} \frac{dA^{(a)}_{1,0}}{\sqrt{2\pi}}.
\end{align}
The rest of the computation proceeds as before, and the final result is
\begin{align}
	Z^{(s=2,m_{2,0}^2)}_\text{PI}=\frac{i^{-1}}{\text{Vol}(G_{1,0})_\text{PI}}\frac{\det'_{-1} |-\nabla_{(0)}^2-(d+1)|^{1/2}}{\det\nolimits'_{-1} \left(-\nabla_{(2)}^2+d+1\right)^{1/2}}.
\end{align}

\subsection{General PM fields}

We now provide a prescription to obtain the path integral expression for a general spin-$s$ depth-$t$ field. First, take the spin-$s$ path integral \eqref{massive general} with generic mass and take the limit $m^2\to m_{s,t}^2$ while omitting the $(t-1,s)$ modes:
\begin{align}
	Z^{(s,m^2\to m_{s,t}^2)}_\text{PI}\to i^{\sum_{m=-1}^{t-2}D^{d+2}_{m,s}}\det\nolimits'_{-1} |-\nabla_{(s)}^2-\lambda_{t-1,s}^2|^{-1/2}
\end{align}
where we have used \eqref{eigen PM}. The phases appear because the mode with $n=-1,0,\cdots, t-2$ becomes negative. Then we exchange $s$ and $t$ and flip $i\to -i$ to obtain another expression
\begin{align}
	Z^{(t,m^2\to m_{t,s}^2)}_\text{PI}\to i^{-\sum_{m=-1}^{s-2}D^{d+2}_{m,t}}\det\nolimits'_{-1} |-\nabla_{(t)}^2-\lambda_{s-1,t}^2|^{-1/2}.
\end{align}
We propose that the final result is simply given by the ratio between these two expressions, divided by a group volume factor:
\begin{align}\label{general s t}
	Z^{(s,m^2= m_{s,t}^2)}_\text{PI}=\frac{i^{\sum_{m=-1}^{t-2}D^{d+2}_{m,s}+\sum_{m=-1}^{s-2}D^{d+2}_{m,t}}}{\text{Vol}(G_{s-1,t})_\text{PI}}\frac{\det\nolimits'_{-1} \left|\frac{-\nabla_{(t)}^2-\lambda_{s-1,t}^2}{M^2}\right|^{1/2}}{\det\nolimits'_{-1} \left|\frac{-\nabla_{(s)}^2-\lambda_{t-1,s}^2}{M^2}\right|^{1/2}}
\end{align}
where 
\begin{align}
	\text{Vol}(G_{s-1,t})_\text{PI} \equiv \int\prod_{a=1}^{D^{d+2}_{s-1,t}} \frac{M^2}{\sqrt{2\pi}} dA_{s-1,t}
\end{align}
Note that we have restored the dimensionful parameter $M$. One can easily verify that \eqref{general s t} reduces to the massless case when $t=s-1$ and the spin-2 depth-0 case when $s=2,t=0$. The division by $Z^{(t,m^2\to m_{t,s}^2)}_\text{PI}$ can be thought of as the decoupling of the spin-$t$ level-$(s-1-t)$ shift-symmetric field from the massive spin-$s$ field as we take $m^2\to m_{s,t}^2$. Note that the ratio of determinants (without the extension to $n=-1$ modes) in \eqref{general s t} and the relations between PM and conformal higher spin partition functions were first discussed in \cite{Tseytlin:2013jya} for $S^4$ and \cite{Tseytlin:2013fca} for $S^6$. 


As we stressed repeatedly, the determination of the group volume factor $\text{Vol}(G_{s-1,t})_\text{PI}$ requires knowledge of the interactions of the parent theory. In the current case, a natural class of parent theories would be the PM generalizations of higher spin theories \cite{Brust:2016zns}, which include a tower of PM gauge fields and a finite number of massive fields. These theories gauge the PM algebras studied in \cite{Joung:2015jza} and are holographic duals to $\Box^k$ CFTs \cite{Brust:2016gjy}. Their 1-loop path integrals would take the form
\begin{align}\label{PM HS PI}
	Z^\text{PM HS}_\text{PI}=\frac{i^P}{\text{Vol}(G)_\text{PI}}\prod_{s,t}\frac{\det\nolimits'_{-1} \left|\frac{-\nabla_{(t)}^2-\lambda_{s-1,t}^2}{M^2}\right|^{1/2}}{\det\nolimits'_{-1} \left|\frac{-\nabla_{(s)}^2-\lambda_{t-1,s}^2}{M^2}\right|^{1/2}}
\end{align}
where 
\begin{align}
	P=\sum_{s,t}\left(\sum_{m=-1}^{t-2}D^{d+2}_{m,s}+\sum_{m=-1}^{s-2}D^{d+2}_{m,t}\right),\quad \text{Vol}(G)_\text{PI}=\prod_{s,t} \text{Vol}(G_{s-1,t})_\text{PI}
\end{align}
There will be analogous problem of relating $\text{Vol}(G)_\text{PI}$ to a canonical volume $\text{Vol}(G)_\text{can}$ (and making sense of the volume itself) as in the massless case, which will give us the dependence on the Newton's constant $G_N$. If we demand $\log Z^\text{PM HS}_\text{PI}$ to be consistent with a universal form as in the massless case \cite{Anninos:2020hfj}, we should take
\begin{align}\label{PM HS blinear}
	\text{Vol($G$)}_\text{PI}=\text{Vol($G$)}_\text{can}\prod_{s,t} \left( \frac{\text{Vol}(S^{d-1})}{8\pi G_N}\frac{M^4}{(d+2s-2)(d+2t-2)}\right)^{\frac{D^{d+2}_{s-1,t}}{2}}
\end{align}
so that the factor $(d+2s-2)(d+2t-2)$ gets nicely canceled upon evaluating the character integrals for the determinants. To verify this, one has to repeat the analysis of \cite{Joung:2013nma} and appendix \ref{cubic} and express the PM HS invariant bilinear form in terms of the bilinear form induced by the path integral measure. Provided that \eqref{PM HS blinear} is valid, we note that except the phase and $\text{Vol($G$)}_\text{can}$, the expression \eqref{PM HS PI} becomes the inverse of itself upon exchanging $s$ and $t$. We leave the validation of \eqref{general s t}, \eqref{PM HS blinear} and the implication of the suggestive $s\leftrightarrow t$ symmetry for future work.


\section{Discussion and outlooks}\label{conclusion}

In this work, we derive the determinant expressions of the 1-loop path integrals for massive, shift-symmetric and (partially) massless symmetric tensor fields on $S^{d+1}$. 

At the technical level, we have clarified subtleties arising from normalizable zero modes or negative modes on the sphere, and have made broad generalizations of all known results. One natural generalization of this work is to study path integrals involving fermionic massive, shift-symmetric and (partially) massless gauge fields. For instance, the free actions for massless fermionic fields are presented in \cite{Fang:1978wz}. Since fermionic fields are Grassman-valued, no Wick rotation is needed to make the path integral convergent. However, there is still a group volume factor corresponding to trivial fermionic gauge transformations, whose physical interpretations are more obscure than their bosonic counterparts, because the Grassman integrals are formally zero. Perhaps we need to combine bosonic and fermionic higher spin fields into a supersymmetric HS theory \cite{Sezgin:2012ag} so that we can make sense of the super-higher-spin group volume.

At the conceptual level, our results and their implications on de Sitter thermodynamics were discussed extensively in \cite{Anninos:2020hfj}. However, the interpretations of a number of features of 1-loop sphere partition functions require further investigations: First is the Polchinski's phase. While we generalize the original massless spin-2 result to other classes of fields, their physical interpretations remain elusive. One is tempted to say perhaps these phases indicate non-unitarity. While this seems to be natural for massive fields with masses outside the unitary bounds (including shift-symmetric fields), the phases are present for PM fields which are perfectly unitary irreducible representations. Without other physical inputs, it is not clear whether we should ignore or retain these phases. However, we stress that these phases deserve our attentions. One reason is that perhaps a better understanding of these phases will lead us to the correct statistical interpretation of the Euclidean path integral.\footnote{For example, one might guess that these $i$'s are precisely the $i$'s present in the inverse Laplace transform to extract microcanonical entropies from the partition function.} Another reason is that $S^{d+1}$ is only one of the many saddle points of the Euclidean gravitational path integral with a positive cosmological constant. If one considers other saddle points such as $S^2\times S^{d-1}$, since they have different amount of symmetries, after Wick rotating the conformal modes there will be \textbf{relative} phases between different saddle points. 

The second remaining mystery is the residual group volume factor present for PM gauge fields and shift-symmetric fields. Such a factor is present for a manifestly local path integral and depends on the non-linear completion of the theory. Higher spin groups are typically infinite-dimensional and there is an issue of making sense of the group volume. The group volume may be more well-defined in theories gauging finite dimensional higher spin algebras studied in \cite{Joung:2015jza}. 


The context in which both subtleties of phases and group volume are sharpest is in $d+1=3$ dimension \cite{Anninos:2020hfj}. In this case one can check that for any PM fields, the determinants for the on-shell kinetic operator and the ghost operator cancel completely, so that the group volume and phases are the \textbf{only} non-trivial contributions to the 1-loop path integral. Also, on $S^3$ there is an alternative formulation of massless HS gravity as a $SU(N)\times SU(N)$ Chern-Simons theory. As noted in \cite{Anninos:2020hfj}, one finds that their 1-loop results agree only if we identify the residual group volume with the $SU(N)\times SU(N)$ HS group volume, further supporting the claim that this factor depends on the interactions of the full theory. Also, the phases will match exactly for odd framing. 

Finally, as mentioned at the end of the introduction, it would be interesting to apply our results to other contexts such as such as string theory, Chern-Simons theory, supersymmetry, AdS/CFT correspondence, conformal field theory, as well as entanglement entropy in quantum field theories. We leave these to future work.

\section*{Acknowledgments} 


I am grateful to Dionysios Anninos, Frederik Denef and Zimo Sun for numerous stimulating discussions, at different stages of this research project. I also thank Manvir Grewal and Klaas Parmentier for their precious comments. Finally, I appreciate the reviewer whose suggestions helped improve and clarify this manuscript. AL  was supported in part by the U.S. Department of Energy grant de-sc0011941.


\appendix


\section{Conventions and definitions}\label{convention}

\paragraph{Symmetrization} We symmetrize a rank-$s$ tensor $\phi_{\mu_1 \cdots \mu_s}$ by summing all the permutations followed by a division by $s!$. That is
\begin{align}
	\phi_{(\mu_1 \cdots \mu_s)}=\frac{1}{s!}\sum_{\sigma \text{:perm}} \phi_{\mu_{\sigma(1)} \cdots \mu_{\sigma(s)}}
\end{align}

\paragraph{Shorthand notations} Throughout this paper we denote
\begin{align}
	\int_{S^{d+1}} \equiv \int_{S^{d+1}} d^{d+1}x \sqrt{g}.
\end{align}
When dealing with a rank-$s$ totally symmetric tensor, we sometimes use the notations:
\begin{align}
	A_{(s)}\equiv&A_{\mu_1 \cdots \mu_s}\\
	g^k \nabla^{(n-2k)}A_{(s)}\equiv&g_{(\mu_1 \mu_2}\cdots g_{\mu_{2k-1} \mu_{2k}} \nabla_{\mu_{2k+1}}\cdots \nabla_{\mu_n}A_{\mu_{n+1} \cdots \mu_{s+n})}\\
	\nabla \cdot A_{(s)}\equiv&\nabla^\lambda A_{\mu_1 \cdots \mu_{s-1} \lambda}\\
	\text{Tr}A_{(s)}\equiv& g^{\lambda \rho}A_{\lambda \rho \mu_1 \cdots \mu_{s-2}}
\end{align}
For two spin-$s$ fields $\psi_{(s)}$ and $\psi'_{(s)}$, we define the inner product
\begin{align}\label{inner}
	( \psi_{(s)} ,\psi'_{(s)}) \equiv \int_{S^{d+1}} \psi^{\mu_1 \cdots \mu_s}\psi'_{\mu_1 \cdots \mu_s}.
\end{align}

\paragraph{Path integral measure}

Path integrals for a spin-$s$ bosonic field $\phi_{(s)}$ take the form 
\begin{align}
	\int \mathcal{D}\phi_{(s)} e^{-\frac{1}{2\mathrm{g}^2}(\phi_{(s)},-\mathcal{Q}\phi_{(s)})}
\end{align}
where $\mathcal{Q}$ is a Laplace type operator. The measure $\mathcal{D}\phi_{(s)}$ is defined as follows. Suppose $\phi_{(s)}$ has mass dimension $\frac{d-2p}{2}$ and an expansion in terms of orthonormal modes $f_{n,(s)}$, i.e.
\begin{align}
	\phi_{(s)} = \sum_n a_{n,s} f_{n,(s)}, \quad ( f_{n,(s)}, f_{m,(s)}) =\delta_{nm}.
\end{align}
We define the path integration measure for $\phi$ to be 
\begin{align}\label{EPImeasure}
	\mathcal{D}\phi_{(s)} \equiv \prod_n \frac{M^p}{\sqrt{2\pi}\mathrm{g}}d a_{n,s}.
\end{align}
Here are some comments:
\begin{itemize}
	\item $M$ is a parameter with mass dimension 1, and the power $p$ is determined by dimension analysis so that the partition function remains dimensionless. In most of this paper we will set $M=1$ and restore it by dimension analysis when necessary.

	\item  The factors of $\sqrt{2\pi}\mathrm{g}$ are inserted such that the path integration results in a determinant without any extra factor other than the dimensionful parameter $M$:
	\begin{align}
		\int \mathcal{D}\phi\, e^{-S[\phi]} = \det(-\frac{\mathcal{Q}}{M^{2p}})^{-1/2} .
	\end{align}
	
	\item The multiplication of the factor $ \frac{M^p}{\sqrt{2\pi}\mathrm{g}}$ only affects UV divergent terms of the resulting free energy and thus can be absorbed into the bare couplings of the local curvature densities. This can be seen as follows. In heat kernel regularization, the path integral is expressed as an integral transform
	\begin{align}
		\log Z_\text{PI} =\int_0^\infty \frac{d\tau}{2\tau}e^{-\frac{\epsilon^2}{4\tau}}\text{Tr}\,e^{-D \tau} 
	\end{align}
	of the trace heat kernel for an unconstrained differential operator $D$. Here for concreteness we have chosen a specific UV regulator $e^{-\frac{\epsilon^2}{4\tau}}$. The result takes the form
	\begin{align}
		\log Z_\text{PI} =\frac{1}{2}\zeta'(0)+\alpha_{d+1}\log (\frac{2}{e^{\gamma_E} \epsilon})+\frac{1}{2}\sum_{k=0}\alpha_k \Gamma\left(\frac{d+1-k}{2}\right)\left(\frac{2}{\epsilon}\right)^{d+1-k}.
	\end{align}
	Here $\zeta(z)$ is the spectral zeta function for the operator $D$. The heat kernel coefficients $\alpha_i$ are given by integrals of local curvature densities (see for example \cite{Vassilevich:2003xt} for explicit formulas). The term $\alpha_{d+1}$ is present only for odd $d$. Now, the multiplication by a local infinite constant is equivalent to rescaling the differential operator by a constant, 
	\begin{align}
		\log Z_\text{PI} \to \log Z'_\text{PI} =\int_0^\infty \frac{d\tau}{2\tau}e^{-\frac{\epsilon^2}{4\tau}}\text{Tr}\,e^{-\tau \frac{(-\nabla^2+\sigma)}{\mathrm{g}}} =\int_0^\infty \frac{d\tau}{2\tau}e^{-\frac{\epsilon^2}{4\mathrm{g}\tau}}\text{Tr}\,e^{-\tau (-\nabla^2+\sigma)} ,
	\end{align}
	which alters only the divergent terms as $\epsilon\to 0$.

	\item With these conventions we also see that the field $\phi$ satisfies the normalization condition
	\begin{align}
		\int \mathcal{D}\phi \,e^{-\frac{1}{2\mathrm{g}^2}(\phi,\phi)}=1.
	\end{align}
	
	\item We can think of the measure \eqref{EPImeasure} as putting the following metric on the field space
	\begin{align}\label{PI metric}
		ds^2=\frac{M^{2p}}{2\pi \mathrm{g}^2}\int_{S^{d+1}} (\delta\phi)^2=\frac{M^{2p}}{2\pi \mathrm{g}^2}\sum_n d a_n^2
	\end{align}

\end{itemize}

\paragraph{Commutator}

In our convention the commutator of two covariant derivatives acts on a totally symmetric rank-$s$ tensor as
\begin{align}
	[\nabla_\mu, \nabla_\nu]\phi^{\rho_1 \cdots \rho_s} =\sum_{j=1}^s R\indices{^{\rho_j}_\lambda_\mu_\nu}\phi^{\rho_1 \cdots \hat{\rho}_j \cdots \rho_s} ,\quad R\indices{_\lambda_\rho_\mu_\nu}=\frac{g_{\lambda \mu}g_{\rho\nu}-g_{\lambda \nu}g_{\rho\mu}}{\ell^2}.
\end{align}
where $\hat{\rho}_j$ means that $\rho_j$ is excluded. $\ell$ is the radius of the sphere and will be set to 1 for most of this paper.

%
%


\section{Symmetric transverse traceless Laplacians and symmetric tensor spherical harmonics on $S^{d+1}$}\label{STSH}

Here we collect some useful facts from \cite{Rubin:1983be} and \cite{Higuchi:1986wu} about spin-$s$ symmetric transverse traceless (STT) Laplacians and symmetric tensor spherical harmonics (STSH) on $S^{d+1}$.

\subsection*{Definition, eigenvalues and degeneracies}

STSHs $f_{n,(s)}\equiv f_{n,\mu_1 \cdots \mu_s}$ are labeled by their spin $s$ and angular momentum number $n\geq s$. These are the STT eigenfunctions of the STT Laplacian $-\nabla_{(s)}^2$ on $S^{d+1}$
\begin{align}\label{STSH eq}
	-\nabla_{(s)}^2f_{n,(s)}=&\lambda_{n,s} f_{n,(s)}, \quad \nabla \cdot f_{n,(s)}=0, \quad \text{Tr}f_{n,(s)}=0 
\end{align}
with eigenvalues and degeneracies
\begin{align}\label{eq:data}
	\lambda_{n,s}=&n(n+d)-s, \quad n\geq s\\
	D^{d+2}_{n,s}=& g_{s}\frac{(n-s+1)(n+s+d-1)(2n+d)(n+d-2)!}{d!(n+1)!}\\
	g_{s} = &\frac{(2s+d-2)(s+d-3)!}{(d-2)!s!}.
\end{align}
These furnish $SO(d+2)$ irreducible representations corresponding to two-row Young diagrams with $n$ boxes in the first row and $s$ boxes in the second row. We sometimes call them $(n,s)$ modes in the paper. We normalize them with respect to \eqref{inner}, i.e.
\begin{align}
	(f_{n,(s)},f_{m,(s)})=\delta_{nm} 
\end{align}
When we use a double labeling such as $f_{n,(s)}$ for the spin-$s$ STSHs or $\lambda_{n,s}$ for its eigenvalues, the $n$ automatically labels the spectrum of $-\nabla_{(s)}^2$. Also, when we write $\sum_{n}$ or $\prod_{n}$, there is an implied sum or product over degenerate spin-$s$ STSHs with the same label $n$. 

\paragraph{Killing tensors}

A spin-$s$ Killing tensor (KT) $\epsilon_{(s)}$ is a totally symmetric traceless tensor satisfying the Killing equation
\begin{align}
	\nabla_{(\mu_1}\epsilon_{\mu_2\cdots \mu_{s+1})}=0.
\end{align}
Taking the trace of this equation shows that they are divergenceless, while taking the divergence we recover \eqref{STSH eq} with $n=s$ and thus they are in fact spanned by the $(s,s)$ modes.

\subsection*{Induced spin-$s$ symmetric traceless spherical harmonics}

Given a STSH $f_{n,(s)}$, one can construct the $m$-th induced symmetric traceless tensors
\begin{align}
	T_{n, (s+m)}^{(s)} = \nabla_{(\mu_1}\cdots \nabla_{\mu_{m}}f_{n, \mu_{m+1}\cdots \mu_{m+s})}-\text{trace terms},
\end{align}
where the subtraction of trace terms is such that the expression is traceless. From its definition, it is clear that $T_{n, (s)}^{(s)} =f_{n, (s)}$. There are two important facts to note:
\begin{enumerate}
	\item  $T_{n,(s)}^{(m)}$ satisfy an orthogonality condition under the inner product (\ref{inner}).
	
	\item  $T_{n,(s)}^{(m)}$ vanishes identically for $s>n$.
\end{enumerate}
The more familiar lower spin examples include the orthonormal modes for the longitudinal part of a vector field
\begin{align*}
	T_{n, \mu }^{(0)}=&\nabla_\mu f_{n}
\end{align*}
or the orthonormal modes for the symmetric traceless part of a spin-2 tensor constructed from scalar spherical harmonics
\begin{align*}
	T_{n, \mu \nu}^{(0)}=&\nabla_{\mu}\nabla_{\nu}f_{n}+\frac{\lambda_{n,0}}{d+1}g_{\mu\nu} f_n.
\end{align*}
We might use the notation $(n,s)$ to refer to a spin-$(s+m)$ symmetric traceless spherical harmonics $T_{n, (s+m)}^{(s)} $ induced from the STSH $f_{n,(s)}$ when the context is clear.

\paragraph{Mode expansions for symmetric  traceless tensors}

In general, a spin-$s$ symmetric traceless (not necessarily transverse) field $V_{(s)}$ on $S^{d+1}$ has the mode expansion
\begin{align}
	V_{(s)} = \sum_{m=0}^s \sum_{n=s}^\infty A_{n,m} \hat{T}_{n, (s)}^{(m)},
\end{align}
where $\hat{T}_{n, (s)}^{(m)}$ is the normalized version of $T_{n, (s)}^{(m)}$, i.e.
\begin{align}
	\hat{T}_{n, (s)}^{(m)}\equiv \frac{T_{n, (s)}^{(m)}}{||T_{n, (s)}^{(m)}||}
\end{align}
where the norm $||\cdot,\cdot|| \equiv \sqrt{(\cdot,\cdot)}$ is defined with respect to \eqref{inner}.

\paragraph{Useful identities}  In this work we make use of the following identities for $T_{n,(s)}^{(m)}$
\begin{align}
	-\nabla^2 T_{n,(s)}^{(m)}&=a_{s,n}^{(m)} T_{n,(s)}^{(m)}\label{eq:Tidentity1}\\
	\nabla \cdot T_{n,(s)}^{(m)}&=b_{s,n}^{(m)} T_{n,(s-1)}^{(m)}\label{eq:Tidentity2}
\end{align}
where
\begin{align}
	a_{s,n}^{(m)} &=\lambda_{n,m} -(s-m)(s +m+d-1) \label{adef}\\
	b_{s,n}^{(m)} &=\frac{(s-m)(d+s+m-2)}{2} (\lambda_{s-1,s}- \lambda_{n,s}).\label{bdef}
\end{align}

\paragraph{Norms} For our purpose, we do not need to know the norm of $T_{n,(s)}^{(m)}$ (with respect to \eqref{inner}), but we need the relative normalizations between $T_{n,(s)}^{(m)}$, $T_{n,(s-1)}^{(m)}$ and $\nabla \cdot T_{n,(s)}^{(m)}$, which can be easily computed:
\begin{align}
	( \nabla\cdot T_{n,(s)}^{(m)},\nabla \cdot T_{n,(s)}^{(m)} )
	=(b_{s,n}^{(m)})^2  ( T_{n,(s-1)}^{(m)},T_{n,(s-1)}^{(m)} ) = -b_{s,n}^{(m)}( T_{n,(s)}^{(m)},T_{n,(s)}^{(m)} ).\label{eq:divTNormidentity1}
\end{align}

\paragraph{Conformal Killing tensors}

A spin-$s$ conformal Killing tensor (CKT) $\epsilon_{(s)}$ is a totally symmetric traceless tensor satisfying the conformal Killing equation
\begin{align}
	\nabla_{(\mu_1}\epsilon_{\mu_2\cdots \mu_{s+1})}-\frac{s}{d+2s-1}g_{(\mu_1\mu_2}\nabla^\lambda \epsilon_{\mu_3\cdots \mu_{s+1})\lambda}=0.
\end{align}
The solution space to this equation is spanned by $T_{s,(s)}^{(m)}$ with $m=0,1,\cdots ,s$. Notice that the modes $T_{s,(s)}^{(s)}=f_{s,{(s)}}$ correspond to spin-$s$ KTs.


\section{Higher spin invariant bilinear form}\label{cubic}

In this appendix, we relate the HS invariant bilinear form in \cite{Joung:2013nma} to the one induced by our path integral measure.

\subsection*{The Noether approach}\label{Noether}

Suppose we have a quadratic action $S^{(2)}$ of a collection of fields $\varphi$ that is invariant under the linear gauge symmetries $\delta_\xi^{(0)}\varphi$, which we want to deform into an interacting action
\begin{align}
	S=S^{(2)}+S^{(3)}+S^{(4)}+\cdots
\end{align}
invariant under the non-linear gauge symmetries
\begin{align}\label{nl gauge trans}
	\delta_\xi \varphi=\delta_\xi^{(0)}\varphi+\delta_\xi^{(1)}\varphi+\delta_\xi^{(2)}\varphi+\cdots .
\end{align}
Here the superscript $(n)$ denotes the power in fields (or coupling constants). Requiring full gauge invariance, i.e.
\begin{align}
	\delta_\xi S=0,
\end{align}
we have a system of equations relating deformations and the gauge transformations at particular orders:
\begin{gather}
	\delta_\xi^{(0)} S^{(2)} =0 \nonumber \\
	\delta_\xi^{(0)} S^{(3)} +\delta_\xi^{(1)} S^{(2)} =0 \nonumber \\
	\delta_\xi^{(0)} S^{(4)} +\delta_\xi^{(1)} S^{(3)} +\delta_\xi^{(2)} S^{(2)} =0 \\
	\cdots  \nonumber
\end{gather}
This can be solved as follows:
\begin{enumerate}
	\item We solve the second equation on the solutions of the first equation $\delta S^{(2)}=0$ to infer the cubic interaction $S^{(3)}$.
	
	\item From this we can infer $\delta_\xi^{(1)}$ by solving the second equation again without imposing the first equation $\delta S^{(2)}=0$.
	
	\item Proceed in a similar fashion for all higher order $S^{(n\geq 3)}$ and the field-dependent part of the gauge transformations $\delta_\xi^{(n\geq1)}$. That is, we solve for $S^{(n)}$ using by the $(n-1)$-th constraint with the $(n-1)$-th order  equation of motion imposed, and then for the deformation $\delta_\xi^{(n-1)}$ by solving the same equation without imposing equations of motion.
\end{enumerate}

\paragraph{Local gauge algebra} The full non-linear gauge transformations \eqref{nl gauge trans} are required to form an (open) algebra 
\begin{align}\label{full nl algebra}
	\delta_{\xi_1}\delta_{\xi_2}-\delta_{\xi_2}\delta_{\xi_1} = \delta_{[[\xi_1,\xi_2]]} + \text{(on-shell trivial)}
\end{align}
where (as illustrated in the Yang Mills and Einstein gravity case) the precise form of the bracket $[[\cdot,\cdot]]$ depends on how gauge transformations act on $\varphi$. In particular, it can be field dependent and can be expanded as
\begin{align}
	[[\cdot,\cdot]]=[[\cdot,\cdot]]^{(0)}+[[\cdot,\cdot]]^{(1)}+[[\cdot,\cdot]]^{(2)}+\cdots.
\end{align}
The full algebra \eqref{full nl algebra} can then be perturbatively expanded in powers of fields.

\paragraph{Global symmetry algebra} We are interested in the global symmetry algebra, the subalgebra of the full local gauge algebra satisfying 
\begin{align}
	\delta^{(0)}=0.
\end{align}
To determine this, it suffices to consider the lowest order:
\begin{align}
	\delta^{(1)}_{\xi_1}\delta^{(0)}_{\xi_2}-\delta^{(1)}_{\xi_2}\delta^{(0)}_{\xi_1} = \delta^{(0)}_{[[\xi_1,\xi_2]]^{(0)}}.
\end{align}
To summarize, the idea is that once the cubic interaction $S^{(3)}$ is determined, we can deduce the deformation of the gauge symmetry $\delta_\xi^{(1)}\varphi$ and the gauge algebra $[[\xi_1,\xi_2]]^{(0)}$:
\begin{align}
	S^{(3)} \implies \delta_\xi^{(1)}\varphi \implies [[\xi_1,\xi_2]]^{(0)},
\end{align}
which then completely fixes the global symmetry algebra and the invariant bilinear form on the algebra (up to an overall normalization). In section \ref{massless spin1} and \ref{massless spin2} we see how it works for Yang-Mills and Einstein theories. Following a similar line of reasoning, the global HS algebra and the HS invariant form has been determined in \cite{Joung:2013nma} for massless higher spin gauge theories. To correctly apply their results in our setting, we are going to make suitable identifications carefully.

\subsection*{Embedding space formalism}

The relevant results in \cite{Joung:2013nma} are expressed in the embedding space formalism. The starting point is to realize $S^{d+1}$ as a $(d+1)$-dimensional hypersurface embedded in an ambient Euclidean space $\mathbb{R}^{d+2}$:
\begin{align}
	X^{2}=(X^1)^2+\cdots +(X^{d+2})^2=\ell_{S^{d+1}}^2
\end{align}
with $l_{S^{d+1}}$ being the radius of the sphere. Symmetric spin-$s$ fields $\phi_{\mu_1 \cdots \mu_s}(x)$ intrinsic to this submanifold are described by an ambient avatar $\Phi_{I_1 \cdots I_s}(X)$ subject to homogeneity and tangentiality constraints
\begin{align}\label{homo tan}
	(X\cdot \partial_X-U\cdot \partial_U +2 +\mu) \Phi(X,U)=0,\quad X\cdot \partial_U \Phi(X,U)=0,
\end{align}
where we have packaged all the $\Phi_{(s)}(X)$ into a generating function
\begin{align}
	\Phi(X,U) =\sum_s \frac{1}{s!}\Phi_{I_1 \cdots I_s}(X) U^{I_1}\cdots U^{I_s}
\end{align}
with an ambient auxiliary vector $U^A$. The homogeneity degree $\mu$ in \eqref{homo tan} is related to the mass of the field. The massless case of interest corresponds to $\mu=0$, in which case we have a gauge symmetry
\begin{align}\label{embed gauge sym}
	\delta_E \Phi (U) = U \cdot \partial_X E(X,U)+O(\Phi)
\end{align}
where the field-dependent part is to be determined by the cubic couplings. The gauge parameter\footnote{There is an extra factor of $\frac{1}{\sqrt{s}}$ compared to \cite{Joung:2013nma}, so that we can identify $E_{I_1 \cdots I_{s-1}}(X)=\Lambda_{I_1 \cdots I_{s-1}}(X)$, with $\Lambda_{I_1 \cdots I_{s-1}}(X)$ being the embedding space representative of $\Lambda_{\mu_1 \cdots \mu_{s-1}}(x)$.}
\begin{align}\label{gauge gen}
	E(X,U)=\sum_s \frac{\sqrt{s}}{s!}E_{I_1 \cdots I_{s-1}}(X) U^{I_1}\cdots U^{I_{s-1}}
\end{align}
satisfies the homogeneity and tangentiality conditions to be consistent with \eqref{homo tan}: 
\begin{align}\label{homo tan E}
	(X\cdot \partial_X-U\cdot \partial_U ) E=0,\quad X\cdot \partial_U E=0.
\end{align}
In this framework, the quadratic action invariant under the linear gauge symmetries \eqref{embed gauge sym} is given by
\begin{align}
	S^{(2)}\stackrel{\text{TT}}{=}-\frac{1}{2}\int_{S^{d+1}} e^{\partial_{U_1}\cdot \partial_{U_2}}\Phi (U_1)\partial_X^2\Phi (U_2)\bigg|_{U_i=0},
\end{align}
where the notation $\stackrel{\text{TT}}{=}$ means equivalence up to trace and divergence terms.
and we are going to construct the cubic vertices following the program described in appendix \ref{Noether}. Note that this normalization is equivalent to choosing 
\begin{align}\label{gs iden}
	\mathrm{g}_s^2 = s!
\end{align}
in \eqref{eq:opaction}.

\subsection*{Killing tensors and global HS algebra}

\paragraph{Killing Tensors}

Global HS symmetries are generated by traceless gauge parameters satisfying the Killing equation
\begin{align}
	U\cdot \partial_X \bar{E}(X,U)=0,\quad \partial_U^2 \bar{E}(X,U)=0.
\end{align}
Together with the homogeneity and tangentiality conditions \eqref{homo tan E} on the gauge parameter, one can also conclude that the Killing tensors satisfy
\begin{align}
	\partial_U\cdot \partial_X \bar{E}(X,U)=0,\quad \partial_X^2 \bar{E}(X,U)=0.
\end{align}
It is straightforward to write down the general solution to these equations:
\begin{gather}
	\bar{E}(X,U)=\sum_r \frac{1}{\sqrt{r+1}}\bar{E}^{(r+1)}(X,U)\nonumber\\
	\bar{E}^{(r+1)}(X,U) = \frac{1}{r!}\bar{E}^{(r+1)}_{I_1 \cdots I_{r}}(X)U^{I_1}\cdots U^{I_{r}}= \frac{1}{(r!)^2}\bar{E}_{I_1 J_1, \cdots,I_r J_r} X^{[I_1}U^{J_1]}\cdots  X^{[I_r}U^{J_r]}.
\end{gather}
The HS generators are the duals of the parameters $\bar{E}_{I_1 J_1, \cdots,I_r J_r}$, which due to the complete tracelessness of the latter are defined as equivalence classes
\begin{align}
	T^{I_1\cdots I_r ,J_1 \cdots J_r}=X^{[I_1}U^{J_1]}\cdots  X^{[I_r}U^{J_r]}+ \cdots 
\end{align}
modulo trace terms $X^2, X\cdot U,U^2$ denoted by $\cdots$ in the equation.

\paragraph{Global HS algebra} Following the framework described in section \ref{Noether}, one can determine the full global HS algebra. What is most relevant to us is the bracket for the spin-2 generators (i.e. Killing vectors), which generate the isometry subalgebra $so(d+2)$:\footnote{We omit the superscript $(0)$ because it is the complete bracket for the global $so(d+2)$ algebra.}
\begin{align}\label{HS s2 bracket}
	[[\bar{E}^{(2)},\bar{E}'^{(2)}]]=-\frac{g}{\sqrt{2}}(\bar{E}^I \partial_I \bar{E}'_J-\bar{E}'^I \partial_I \bar{E}_J)U^J,
\end{align}
where $g$ is the coupling constant of the theory, which can be identified with the Newton's constant through
\begin{align}\label{HS coupling iden}
	g^2=32 \pi G_N.
\end{align}
To obtain \eqref{HS s2 bracket}, one can recall the footnote around \eqref{Einstein bracket}, and note that there is an extra factor of $\frac{1}{\sqrt{2}}$ because of the non-canonical normalization due to the identification \eqref{gs iden}. Canonical generators are those satisfying the standard $so(d+2)$ commutation relation under the bracket \eqref{HS s2 bracket}:
\begin{align}
	[[M_{IJ},M_{KL}]]=\eta_{JK}M_{IL}-\eta_{JL}M_{IK}+\eta_{IL}M_{JK}-\eta_{IK}M_{JL}.
\end{align}
One such basis is $M_{IJ}=-\frac{\sqrt{2}}{g}(X_IU_J-X_J U_I)$ with $I,J=1,\cdots,d+2$, with which we will fix the overall normalization of the canonical metric. In general, the higher spin commutators mix Killing tensors with different spins. For example, a commutator of two spin-3 generators is a linear combination of a spin-2 and a spin-4 generator
\begin{align}
	[[ \bar{E}^{(3)},\bar{E}'^{(3)}]]\sim \bar{E}^{(2)}+\bar{E}^{(4)}.
\end{align}
Fortunately, upon the identifications \eqref{gs iden}, the HS invariant bilinear form obtained in \cite{Joung:2013nma} is uniquely related to our path integral metric (up to an overall normalization), and therefore the knowledge of the brackets for all higher spin generators is not needed.

\subsection*{HS invariant bilinear form}

The HS bilinear form takes the general form\footnote{The factor of $\frac{1}{\sqrt{s}}$ came from \eqref{gauge gen}.}
\begin{align}\label{HS bi form}
	\bra{\bar{E}_1}\ket{\bar{E}_2} =& \sum_s \frac{b_s}{s} \frac{(\partial_{U_1}\cdot\partial_{U_2})^{s-1}}{(s-1)!}\frac{(\partial_{X_1}\cdot\partial_{X_2})^{s-1}}{(s-1)!}\bar{E}_1 (X_1,U_1)\bar{E}_2 (X_2,U_2)\bigg|_{X_i=U_i=0}\nonumber\\
	=& \sum_s \frac{b_s}{s!} \frac{(\partial_{X_1}\cdot\partial_{X_2})^{s-1}}{(s-1)!} \bar{E}^{(s)}_1(X_1) \bar{E}^{(s)}_2(X_2)\bigg|_{X_i=0}
\end{align}
where the constant $b_s$ is fixed by requiring the cyclic property
\begin{align}
	\bra{\bar{E}_1}\ket{[[\bar{E}_2,\bar{E}_3]]} =\bra{\bar{E}_2}\ket{[[\bar{E}_3,\bar{E}_1]]}=\bra{\bar{E}_3}\ket{[[\bar{E}_1,\bar{E}_2]]}.
\end{align}
For $AdS_{d+1}$ it was determined to be \cite{Joung:2013nma}\footnote{As noted in \cite{Anninos:2020hfj}, we have corrected what we believe to be a typo in \cite{Joung:2013nma}.}
\begin{align}
	b^{AdS_{d+1}}_s =  \frac{b^{AdS_{d+1}}_2(-\ell_{AdS}^2)^{s-2}\Gamma(\frac{d}{2})}{2^{s-2}\Gamma(\frac{d}{2}+s-2)}=\frac{b^{AdS_{d+1}}_2(-\ell_{AdS}^2)^{s-2} }{d(d+2)\cdots (d+2s-8)(d+2s-6)}
\end{align}
where $b^{AdS_{d+1}}_2$ is an overall $s$-independent normalization constant and we have restored the $AdS$ length $\ell_{AdS}$. Wick rotating this to $S^{d+1}$ mounts to replacing $\ell_{AdS}= i \ell_{S^{d+1}}$ and thus
\begin{align}
	b^{S^{d+1}}_s = \frac{ b^{S^{d+1}}_2(\ell_{S^{d+1}})^{s-2}\Gamma(\frac{d}{2})}{2^{s-2}\Gamma(\frac{d}{2}+s-2)}=\frac{b^{S^{d+1}}_2(\ell_{S^{d+1}}^2)^{s-2} }{d(d+2)\cdots (d+2s-8)(d+2s-6)}.
\end{align}
From now on we set $\ell_{S^{d+1}}=1$.

\paragraph{Relation to path integral metric}

In the current notations, the bilinear form for a particular spin induced by the path integral measure is simply
\begin{align}\label{PI bilin}
	\bra{\bar{E}^{(s)}_1}\ket{\bar{E}^{(s)}_2}_\text{PI}=\frac{(s-1)!}{2\pi \mathrm{g}_s^2}\int_{S^{d+1}}\bar{E}_1^{(s)}(X,\partial_U)\bar{E}_2^{(s)}(X,U).
\end{align}
The HS invariant form \eqref{HS bi form} is a linear combination of these
\begin{align}\label{HS PI comb}
	\bra{\bar{E}_1}\ket{\bar{E}_{2}}=\sum_s B_s \bra{\bar{E}^{(s)}_1}\ket{\bar{E}^{(s)}_2}_\text{PI}.
\end{align}
We want to determine the $s$-dependence of the coefficient $B_s$. To that end we note that the contraction in \eqref{HS bi form} can be written as\footnote{To see this, note that
	\begin{align}
		\int_{\mathbb{R}^{d+2}}e^{-X^2/2}X_{I_1}\cdots X_{I_{s-1}}X_{J_1}\cdots X_{J_{s-1}}=\frac{(2\pi)^{\frac{d+2}{2}}}{2^{s-1}(s-1)!}  \Big( \delta_{I_1 J_1}\cdots \delta_{I_{s-1} J_{s-1}}+\text{perm} \Big).\nonumber
	\end{align}
	Here ``perm'' includes all permutations among $\{ I_1 ,J_1,I_2,J_2,\cdots,I_{s-1}, J_{s-1}\}$. In particular, it includes terms like $\delta_{I_1 I_2} \cdots $, which do not contribute in the inner product \eqref{inner integral} since $\bar{E}_1$ and $\bar{E}_2$ are traceless. Therefore, among the $(2s-2)!$ permutations, only $2^{s-1}((s-1)!)^2$ of them gives non-zero contributions in \eqref{inner integral}. }
\begin{align}\label{inner integral}
	\frac{(\partial_{X_1}\cdot\partial_{X_2})^{s-1}}{(s-1)!} \bar{E}^{(s)}_1(X_1) \bar{E}^{(s)}_2(X_2)\bigg|_{X_i=0}=\frac{\int_{\mathbb{R}^{d+2}}e^{-X^2/2}\bar{E}_1^{(s)}(X,\partial_U)\bar{E}_2^{(s)}(X,U)}{\int_{\mathbb{R}^{d+2}}e^{-X^2/2}}.
\end{align}
Computing the integral on the right hand side in the radial coordinates, we have 
\begin{align}
	\int_{\mathbb{R}^{d+2}}e^{-X^2/2}\bar{E}_1^{(s)}(X,\partial_U)\bar{E}_2^{(s)}(X,U)=2^{s+\frac{d}{2}-1}\Gamma\left(\frac{d}{2}+s\right) \int_{S^{d+1}}\bar{E}_1^{(s)}(X,\partial_U)\bar{E}_2^{(s)}(X,U)
\end{align}
Now, with the identification \eqref{gs iden}, comparing \eqref{HS bi form} with \eqref{HS PI comb}, we conclude
\begin{align}
	B_s \propto (d+2s-2)(d+2s-4)
\end{align}
up to a $s$-independent overall normalization constant.

\paragraph{Canonical isometry generators}

What we have so far is the HS invariant bilinear form up to an overall normalization factor
\begin{align}
	\bra{\bar{E}_1}\ket{\bar{E}_{2}}_\text{can}=C\sum_s (d+2s-2)(d+2s-4)  \bra{\bar{E}^{(s)}_1}\ket{\bar{E}^{(s)}_2}_\text{PI}\, .
\end{align}
We fix $C$ by requiring the canonical isometry generators $M_{IJ}=-\frac{\sqrt{2}}{\mathrm{g}}(X_IU_J-X_J U_I)$ to be unit-normalized. Evaluating
\begin{align}
	1=\bra{M_{12}}\ket{M_{12}}_\text{can}=2 C \,  d(d+2)\bra{M_{12}}\ket{M_{12}}_\text{PI}= \frac{4C}{\mathrm{g}^2}\text{Vol}(S^{d-1}),
\end{align}
we fix
\begin{align}
	C= \frac{8\pi G_N}{\text{Vol}(S^{d-1})}
\end{align}
upon the identification \eqref{HS coupling iden}. To conclude, we have found
\begin{align}
	\bra{\bar{E}_1}\ket{\bar{E}_{2}}_\text{can}=\frac{8\pi G_N}{\text{Vol}(S^{d-1})}\sum_s (d+2s-2)(d+2s-4)  \bra{\bar{E}^{(s)}_1}\ket{\bar{E}^{(s)}_2}_\text{PI}\, ,
\end{align}
which leads to the relation \eqref{HS can vol}.



\bibliographystyle{plain}
\bibliography{references}

\providecommand{\href}[2]{#2}\begingroup\raggedright\begin{thebibliography}{10}

\bibitem{PhysRevD.15.2752}
G.~W. Gibbons and S.~W. Hawking, ``Action integrals and partition functions in
  quantum gravity,'' \href{http://dx.doi.org/10.1103/PhysRevD.15.2752}{{\em
  Phys. Rev. D} {\bfseries 15} (May, 1977) 2752--2756}.
  \url{https://link.aps.org/doi/10.1103/PhysRevD.15.2752}.

\bibitem{Gibbons:1978ji}
G.~W. Gibbons and M.~J. Perry, ``{Quantizing Gravitational Instantons},''
  \href{http://dx.doi.org/10.1016/0550-3213(78)90434-0}{{\em Nucl. Phys. B}
  {\bfseries 146} (1978) 90--108}.

\bibitem{Christensen:1979iy}
S.~M. Christensen and M.~J. Duff, ``{Quantizing Gravity with a Cosmological
  Constant},'' \href{http://dx.doi.org/10.1016/0550-3213(80)90423-X}{{\em Nucl.
  Phys. B} {\bfseries 170} (1980) 480--506}.

\bibitem{Fradkin:1983mq}
E.~S. Fradkin and A.~A. Tseytlin, ``{One Loop Effective Potential in Gauged
  O(4) Supergravity},''
  \href{http://dx.doi.org/10.1016/0550-3213(84)90074-9}{{\em Nucl. Phys. B}
  {\bfseries 234} (1984) 472}.

\bibitem{Allen:1983dg}
B.~Allen, ``{Phase Transitions in de Sitter Space},''
  \href{http://dx.doi.org/10.1016/0550-3213(83)90470-4}{{\em Nucl. Phys. B}
  {\bfseries 226} (1983) 228--252}.

\bibitem{Taylor:1989ua}
T.~R. Taylor and G.~Veneziano, ``{Quantum Gravity at Large Distances and the
  Cosmological Constant},''
  \href{http://dx.doi.org/10.1016/0550-3213(90)90615-K}{{\em Nucl. Phys. B}
  {\bfseries 345} (1990) 210--230}.

\bibitem{GRIFFIN1989295}
P.~A. Griffin and D.~A. Kosower, ``Curved spacetime one-loop gravity in a
  physical gauge,''
  \href{http://dx.doi.org/https://doi.org/10.1016/0370-2693(89)91313-0}{{\em
  Physics Letters B} {\bfseries 233} no.~3, (1989) 295--300}.
  \url{https://www.sciencedirect.com/science/article/pii/0370269389913130}.

\bibitem{Mazur:1989ch}
P.~O. Mazur and E.~Mottola, ``{ABSENCE OF PHASE IN THE SUM OVER SPHERES},''.

\bibitem{Vassilevich:1992rk}
D.~V. Vassilevich, ``{One loop quantum gravity on de Sitter space},''
  \href{http://dx.doi.org/10.1142/S0217751X93000679}{{\em Int. J. Mod. Phys. A}
  {\bfseries 8} (1993) 1637--1652}.

\bibitem{Volkov:2000ih}
M.~S. Volkov and A.~Wipf, ``{Black hole pair creation in de Sitter space: A
  Complete one loop analysis},''
  \href{http://dx.doi.org/10.1016/S0550-3213(00)00287-X}{{\em Nucl. Phys. B}
  {\bfseries 582} (2000) 313--362},
  \href{http://arxiv.org/abs/hep-th/0003081}{{\ttfamily arXiv:hep-th/0003081}}.

\bibitem{Anninos:2020hfj}
D.~Anninos, F.~Denef, Y.~T.~A. Law, and Z.~Sun, ``{Quantum de Sitter horizon
  entropy from quasicanonical bulk, edge, sphere and topological string
  partition functions},'' \href{http://arxiv.org/abs/2009.12464}{{\ttfamily
  arXiv:2009.12464 [hep-th]}}.

\bibitem{Polchinski:1988ua}
J.~Polchinski, ``{The Phase of the Sum Over Spheres},''
  \href{http://dx.doi.org/10.1016/0370-2693(89)90387-0}{{\em Phys. Lett. B}
  {\bfseries 219} (1989) 251--257}.

\bibitem{Donnelly:2013tia}
W.~Donnelly and A.~C. Wall, ``{Unitarity of Maxwell theory on curved spacetimes
  in the covariant formalism},''
  \href{http://dx.doi.org/10.1103/PhysRevD.87.125033}{{\em Phys. Rev. D}
  {\bfseries 87} no.~12, (2013) 125033},
  \href{http://arxiv.org/abs/1303.1885}{{\ttfamily arXiv:1303.1885 [hep-th]}}.

\bibitem{Gibbons:1978ac}
G.~W. Gibbons, S.~W. Hawking, and M.~J. Perry, ``{Path Integrals and the
  Indefiniteness of the Gravitational Action},''
  \href{http://dx.doi.org/10.1016/0550-3213(78)90161-X}{{\em Nucl. Phys. B}
  {\bfseries 138} (1978) 141--150}.

\bibitem{Babelon:1979wd}
O.~Babelon and C.~M. Viallet, ``{The Geometrical Interpretation of the
  {Faddeev-Popov} Determinant},''
  \href{http://dx.doi.org/10.1016/0370-2693(79)90589-6}{{\em Phys. Lett. B}
  {\bfseries 85} (1979) 246--248}.

\bibitem{Mazur:1989by}
P.~O. Mazur and E.~Mottola, ``{The Gravitational Measure, Solution of the
  Conformal Factor Problem and Stability of the Ground State of Quantum
  Gravity},'' \href{http://dx.doi.org/10.1016/0550-3213(90)90268-I}{{\em Nucl.
  Phys. B} {\bfseries 341} (1990) 187--212}.

\bibitem{Bern:1990bh}
Z.~Bern, E.~Mottola, and S.~K. Blau, ``{General covariance of the path integral
  for quantum gravity},''
  \href{http://dx.doi.org/10.1103/PhysRevD.43.1212}{{\em Phys. Rev. D}
  {\bfseries 43} (1991) 1212--1222}.

\bibitem{Giombi:2015haa}
S.~Giombi, I.~R. Klebanov, and G.~Tarnopolsky, ``{Conformal QED$_d$,
  $F$-Theorem and the $\epsilon$ Expansion},''
  \href{http://dx.doi.org/10.1088/1751-8113/49/13/135403}{{\em J. Phys. A}
  {\bfseries 49} no.~13, (2016) 135403},
  \href{http://arxiv.org/abs/1508.06354}{{\ttfamily arXiv:1508.06354
  [hep-th]}}.

\bibitem{Joung:2013nma}
E.~Joung and M.~Taronna, ``{Cubic-interaction-induced deformations of
  higher-spin symmetries},''
  \href{http://dx.doi.org/10.1007/JHEP03(2014)103}{{\em JHEP} {\bfseries 03}
  (2014) 103}, \href{http://arxiv.org/abs/1311.0242}{{\ttfamily arXiv:1311.0242
  [hep-th]}}.

\bibitem{Anninos:2011ui}
D.~Anninos, T.~Hartman, and A.~Strominger, ``{Higher Spin Realization of the
  dS/CFT Correspondence},''
  \href{http://dx.doi.org/10.1088/1361-6382/34/1/015009}{{\em Class. Quant.
  Grav.} {\bfseries 34} no.~1, (2017) 015009},
  \href{http://arxiv.org/abs/1108.5735}{{\ttfamily arXiv:1108.5735 [hep-th]}}.

\bibitem{Anninos:2012ft}
D.~Anninos, F.~Denef, and D.~Harlow, ``{Wave function of
  Vasiliev\textquoteright{}s universe: A few slices thereof},''
  \href{http://dx.doi.org/10.1103/PhysRevD.88.084049}{{\em Phys. Rev. D}
  {\bfseries 88} no.~8, (2013) 084049},
  \href{http://arxiv.org/abs/1207.5517}{{\ttfamily arXiv:1207.5517 [hep-th]}}.

\bibitem{Anninos:2013rza}
D.~Anninos, F.~Denef, G.~Konstantinidis, and E.~Shaghoulian, ``{Higher Spin de
  Sitter Holography from Functional Determinants},''
  \href{http://dx.doi.org/10.1007/JHEP02(2014)007}{{\em JHEP} {\bfseries 02}
  (2014) 007}, \href{http://arxiv.org/abs/1305.6321}{{\ttfamily arXiv:1305.6321
  [hep-th]}}.

\bibitem{Anninos:2017eib}
D.~Anninos, F.~Denef, R.~Monten, and Z.~Sun, ``{Higher Spin de Sitter Hilbert
  Space},'' \href{http://dx.doi.org/10.1007/JHEP10(2019)071}{{\em JHEP}
  {\bfseries 10} (2019) 071}, \href{http://arxiv.org/abs/1711.10037}{{\ttfamily
  arXiv:1711.10037 [hep-th]}}.

\bibitem{Vasiliev:1990en}
M.~A. Vasiliev, ``{Consistent equation for interacting gauge fields of all
  spins in (3+1)-dimensions},''
  \href{http://dx.doi.org/10.1016/0370-2693(90)91400-6}{{\em Phys. Lett. B}
  {\bfseries 243} (1990) 378--382}.

\bibitem{Vasiliev:2003ev}
M.~A. Vasiliev, ``{Nonlinear equations for symmetric massless higher spin
  fields in (A)dS(d)},''
  \href{http://dx.doi.org/10.1016/S0370-2693(03)00872-4}{{\em Phys. Lett. B}
  {\bfseries 567} (2003) 139--151},
  \href{http://arxiv.org/abs/hep-th/0304049}{{\ttfamily arXiv:hep-th/0304049}}.

\bibitem{Bekaert:2005vh}
X.~Bekaert, S.~Cnockaert, C.~Iazeolla, and M.~A. Vasiliev, ``{Nonlinear higher
  spin theories in various dimensions},'' in {\em {1st Solvay Workshop on
  Higher Spin Gauge Theories}}.
\newblock 2004.
\newblock \href{http://arxiv.org/abs/hep-th/0503128}{{\ttfamily
  arXiv:hep-th/0503128}}.

\bibitem{Boulanger:2015ova}
N.~Boulanger, P.~Kessel, E.~D. Skvortsov, and M.~Taronna, ``{Higher spin
  interactions in four-dimensions: Vasiliev versus Fronsdal},''
  \href{http://dx.doi.org/10.1088/1751-8113/49/9/095402}{{\em J. Phys. A}
  {\bfseries 49} no.~9, (2016) 095402},
  \href{http://arxiv.org/abs/1508.04139}{{\ttfamily arXiv:1508.04139
  [hep-th]}}.

\bibitem{Sleight:2017pcz}
C.~Sleight and M.~Taronna, ``{Higher-Spin Gauge Theories and Bulk Locality},''
  \href{http://dx.doi.org/10.1103/PhysRevLett.121.171604}{{\em Phys. Rev.
  Lett.} {\bfseries 121} no.~17, (2018) 171604},
  \href{http://arxiv.org/abs/1704.07859}{{\ttfamily arXiv:1704.07859
  [hep-th]}}.

\bibitem{Gaberdiel:2010ar}
M.~R. Gaberdiel, R.~Gopakumar, and A.~Saha, ``{Quantum $W$-symmetry in
  $AdS_3$},'' \href{http://dx.doi.org/10.1007/JHEP02(2011)004}{{\em JHEP}
  {\bfseries 02} (2011) 004}, \href{http://arxiv.org/abs/1009.6087}{{\ttfamily
  arXiv:1009.6087 [hep-th]}}.

\bibitem{Gupta:2012he}
R.~K. Gupta and S.~Lal, ``{Partition Functions for Higher-Spin theories in
  AdS},'' \href{http://dx.doi.org/10.1007/JHEP07(2012)071}{{\em JHEP}
  {\bfseries 07} (2012) 071}, \href{http://arxiv.org/abs/1205.1130}{{\ttfamily
  arXiv:1205.1130 [hep-th]}}.

\bibitem{Giombi:2013fka}
S.~Giombi and I.~R. Klebanov, ``{One Loop Tests of Higher Spin AdS/CFT},''
  \href{http://dx.doi.org/10.1007/JHEP12(2013)068}{{\em JHEP} {\bfseries 12}
  (2013) 068}, \href{http://arxiv.org/abs/1308.2337}{{\ttfamily arXiv:1308.2337
  [hep-th]}}.

\bibitem{Giombi:2014iua}
S.~Giombi, I.~R. Klebanov, and B.~R. Safdi, ``{Higher Spin AdS$_{d+1}$/CFT$_d$
  at One Loop},'' \href{http://dx.doi.org/10.1103/PhysRevD.89.084004}{{\em
  Phys. Rev. D} {\bfseries 89} no.~8, (2014) 084004},
  \href{http://arxiv.org/abs/1401.0825}{{\ttfamily arXiv:1401.0825 [hep-th]}}.

\bibitem{Giombi:2016pvg}
S.~Giombi, I.~R. Klebanov, and Z.~M. Tan, ``{The ABC of Higher-Spin AdS/CFT},''
  \href{http://dx.doi.org/10.3390/universe4010018}{{\em Universe} {\bfseries 4}
  no.~1, (2018) 18}, \href{http://arxiv.org/abs/1608.07611}{{\ttfamily
  arXiv:1608.07611 [hep-th]}}.

\bibitem{Gunaydin:2016amv}
M.~G\"unaydin, E.~D. Skvortsov, and T.~Tran, ``{Exceptional $F(4)$ higher-spin
  theory in AdS$_{6}$ at one-loop and other tests of duality},''
  \href{http://dx.doi.org/10.1007/JHEP11(2016)168}{{\em JHEP} {\bfseries 11}
  (2016) 168}, \href{http://arxiv.org/abs/1608.07582}{{\ttfamily
  arXiv:1608.07582 [hep-th]}}.

\bibitem{Sleight:2017cax}
C.~Sleight and M.~Taronna, ``{Feynman rules for higher-spin gauge fields on
  AdS$_{d+1}$},'' \href{http://dx.doi.org/10.1007/JHEP01(2018)060}{{\em JHEP}
  {\bfseries 01} (2018) 060}, \href{http://arxiv.org/abs/1708.08668}{{\ttfamily
  arXiv:1708.08668 [hep-th]}}.

\bibitem{Fronsdal:1978rb}
C.~Fronsdal, ``{Massless Fields with Integer Spin},''
  \href{http://dx.doi.org/10.1103/PhysRevD.18.3624}{{\em Phys. Rev. D}
  {\bfseries 18} (1978) 3624}.

\bibitem{Higuchi:1986py}
A.~Higuchi, ``{Forbidden Mass Range for Spin-2 Field Theory in De Sitter
  Space-time},'' \href{http://dx.doi.org/10.1016/0550-3213(87)90691-2}{{\em
  Nucl. Phys. B} {\bfseries 282} (1987) 397--436}.

\bibitem{Basile:2016aen}
T.~Basile, X.~Bekaert, and N.~Boulanger, ``{Mixed-symmetry fields in de Sitter
  space: a group theoretical glance},''
  \href{http://dx.doi.org/10.1007/JHEP05(2017)081}{{\em JHEP} {\bfseries 05}
  (2017) 081}, \href{http://arxiv.org/abs/1612.08166}{{\ttfamily
  arXiv:1612.08166 [hep-th]}}.

\bibitem{Zinoviev:2001dt}
Y.~M. Zinoviev, ``{On massive high spin particles in AdS},''
  \href{http://arxiv.org/abs/hep-th/0108192}{{\ttfamily arXiv:hep-th/0108192}}.

\bibitem{Bonifacio:2018zex}
J.~Bonifacio, K.~Hinterbichler, A.~Joyce, and R.~A. Rosen, ``{Shift Symmetries
  in (Anti) de Sitter Space},''
  \href{http://dx.doi.org/10.1007/JHEP02(2019)178}{{\em JHEP} {\bfseries 02}
  (2019) 178}, \href{http://arxiv.org/abs/1812.08167}{{\ttfamily
  arXiv:1812.08167 [hep-th]}}.

\bibitem{Allen:1985ux}
B.~Allen, ``{Vacuum States in de Sitter Space},''
  \href{http://dx.doi.org/10.1103/PhysRevD.32.3136}{{\em Phys. Rev. D}
  {\bfseries 32} (1985) 3136}.

\bibitem{Allen:1987tz}
B.~Allen and A.~Folacci, ``{The Massless Minimally Coupled Scalar Field in De
  Sitter Space},'' \href{http://dx.doi.org/10.1103/PhysRevD.35.3771}{{\em Phys.
  Rev. D} {\bfseries 35} (1987) 3771}.

\bibitem{Bros:2010wa}
J.~Bros, H.~Epstein, and U.~Moschella, ``{Scalar tachyons in the de Sitter
  universe},'' \href{http://dx.doi.org/10.1007/s11005-010-0406-4}{{\em Lett.
  Math. Phys.} {\bfseries 93} (2010) 203--211},
  \href{http://arxiv.org/abs/1003.1396}{{\ttfamily arXiv:1003.1396 [hep-th]}}.

\bibitem{Epstein:2014jaa}
H.~Epstein and U.~Moschella, ``{de Sitter tachyons and related topics},''
  \href{http://dx.doi.org/10.1007/s00220-015-2308-x}{{\em Commun. Math. Phys.}
  {\bfseries 336} no.~1, (2015) 381--430},
  \href{http://arxiv.org/abs/1403.3319}{{\ttfamily arXiv:1403.3319 [hep-th]}}.

\bibitem{Deser:1983tm}
S.~Deser and R.~I. Nepomechie, ``{Anomalous Propagation of Gauge Fields in
  Conformally Flat Spaces},''
  \href{http://dx.doi.org/10.1016/0370-2693(83)90317-9}{{\em Phys. Lett. B}
  {\bfseries 132} (1983) 321--324}.

\bibitem{DESER1984396}
S.~Deser and R.~I. Nepomechie, ``Gauge invariance versus masslessness in de
  sitter spaces,''
  \href{http://dx.doi.org/https://doi.org/10.1016/0003-4916(84)90156-8}{{\em
  Annals of Physics} {\bfseries 154} no.~2, (1984) 396--420}.
  \url{https://www.sciencedirect.com/science/article/pii/0003491684901568}.

\bibitem{Brink:2000ag}
L.~Brink, R.~R. Metsaev, and M.~A. Vasiliev, ``{How massless are massless
  fields in AdS(d)},''
  \href{http://dx.doi.org/10.1016/S0550-3213(00)00402-8}{{\em Nucl. Phys. B}
  {\bfseries 586} (2000) 183--205},
  \href{http://arxiv.org/abs/hep-th/0005136}{{\ttfamily arXiv:hep-th/0005136}}.

\bibitem{Deser:2001pe}
S.~Deser and A.~Waldron, ``{Gauge invariances and phases of massive higher
  spins in (A)dS},''
  \href{http://dx.doi.org/10.1103/PhysRevLett.87.031601}{{\em Phys. Rev. Lett.}
  {\bfseries 87} (2001) 031601},
  \href{http://arxiv.org/abs/hep-th/0102166}{{\ttfamily arXiv:hep-th/0102166}}.

\bibitem{Deser:2001us}
S.~Deser and A.~Waldron, ``{Partial masslessness of higher spins in (A)dS},''
  \href{http://dx.doi.org/10.1016/S0550-3213(01)00212-7}{{\em Nucl. Phys. B}
  {\bfseries 607} (2001) 577--604},
  \href{http://arxiv.org/abs/hep-th/0103198}{{\ttfamily arXiv:hep-th/0103198}}.

\bibitem{Deser:2001wx}
S.~Deser and A.~Waldron, ``{Stability of massive cosmological gravitons},''
  \href{http://dx.doi.org/10.1016/S0370-2693(01)00523-8}{{\em Phys. Lett. B}
  {\bfseries 508} (2001) 347--353},
  \href{http://arxiv.org/abs/hep-th/0103255}{{\ttfamily arXiv:hep-th/0103255}}.

\bibitem{Deser:2001xr}
S.~Deser and A.~Waldron, ``{Null propagation of partially massless higher spins
  in (A)dS and cosmological constant speculations},''
  \href{http://dx.doi.org/10.1016/S0370-2693(01)00756-0}{{\em Phys. Lett. B}
  {\bfseries 513} (2001) 137--141},
  \href{http://arxiv.org/abs/hep-th/0105181}{{\ttfamily arXiv:hep-th/0105181}}.

\bibitem{Skvortsov:2006at}
E.~D. Skvortsov and M.~A. Vasiliev, ``{Geometric formulation for partially
  massless fields},''
  \href{http://dx.doi.org/10.1016/j.nuclphysb.2006.06.019}{{\em Nucl. Phys. B}
  {\bfseries 756} (2006) 117--147},
  \href{http://arxiv.org/abs/hep-th/0601095}{{\ttfamily arXiv:hep-th/0601095}}.

\bibitem{Hinterbichler:2016fgl}
K.~Hinterbichler and A.~Joyce, ``{Manifest Duality for Partially Massless
  Higher Spins},'' \href{http://dx.doi.org/10.1007/JHEP09(2016)141}{{\em JHEP}
  {\bfseries 09} (2016) 141}, \href{http://arxiv.org/abs/1608.04385}{{\ttfamily
  arXiv:1608.04385 [hep-th]}}.

\bibitem{Tseytlin:2013jya}
A.~A. Tseytlin, ``{On partition function and Weyl anomaly of conformal higher
  spin fields},'' \href{http://dx.doi.org/10.1016/j.nuclphysb.2013.10.009}{{\em
  Nucl. Phys. B} {\bfseries 877} (2013) 598--631},
  \href{http://arxiv.org/abs/1309.0785}{{\ttfamily arXiv:1309.0785 [hep-th]}}.

\bibitem{Tseytlin:2013fca}
A.~A. Tseytlin, ``{Weyl anomaly of conformal higher spins on six-sphere},''
  \href{http://dx.doi.org/10.1016/j.nuclphysb.2013.10.008}{{\em Nucl. Phys. B}
  {\bfseries 877} (2013) 632--646},
  \href{http://arxiv.org/abs/1310.1795}{{\ttfamily arXiv:1310.1795 [hep-th]}}.

\bibitem{Brust:2016zns}
C.~Brust and K.~Hinterbichler, ``{Partially Massless Higher-Spin Theory},''
  \href{http://dx.doi.org/10.1007/JHEP02(2017)086}{{\em JHEP} {\bfseries 02}
  (2017) 086}, \href{http://arxiv.org/abs/1610.08510}{{\ttfamily
  arXiv:1610.08510 [hep-th]}}.

\bibitem{Joung:2015jza}
E.~Joung and K.~Mkrtchyan, ``{Partially-massless higher-spin algebras and their
  finite-dimensional truncations},''
  \href{http://dx.doi.org/10.1007/JHEP01(2016)003}{{\em JHEP} {\bfseries 01}
  (2016) 003}, \href{http://arxiv.org/abs/1508.07332}{{\ttfamily
  arXiv:1508.07332 [hep-th]}}.

\bibitem{Brust:2016gjy}
C.~Brust and K.~Hinterbichler, ``{Free \ensuremath{\square}$^{k}$ scalar
  conformal field theory},''
  \href{http://dx.doi.org/10.1007/JHEP02(2017)066}{{\em JHEP} {\bfseries 02}
  (2017) 066}, \href{http://arxiv.org/abs/1607.07439}{{\ttfamily
  arXiv:1607.07439 [hep-th]}}.

\bibitem{Fang:1978wz}
J.~Fang and C.~Fronsdal, ``{Massless Fields with Half Integral Spin},''
  \href{http://dx.doi.org/10.1103/PhysRevD.18.3630}{{\em Phys. Rev. D}
  {\bfseries 18} (1978) 3630}.

\bibitem{Sezgin:2012ag}
E.~Sezgin and P.~Sundell, ``{Supersymmetric Higher Spin Theories},''
  \href{http://dx.doi.org/10.1088/1751-8113/46/21/214022}{{\em J. Phys. A}
  {\bfseries 46} (2013) 214022},
  \href{http://arxiv.org/abs/1208.6019}{{\ttfamily arXiv:1208.6019 [hep-th]}}.

\bibitem{Vassilevich:2003xt}
D.~V. Vassilevich, ``{Heat kernel expansion: User's manual},''
  \href{http://dx.doi.org/10.1016/j.physrep.2003.09.002}{{\em Phys. Rept.}
  {\bfseries 388} (2003) 279--360},
  \href{http://arxiv.org/abs/hep-th/0306138}{{\ttfamily arXiv:hep-th/0306138}}.

\bibitem{Rubin:1983be}
M.~A. Rubin and C.~R. Ordonez, ``{EIGENVALUES AND DEGENERACIES FOR
  n-DIMENSIONAL TENSOR SPHERICAL HARMONICS},''.

\bibitem{Higuchi:1986wu}
A.~Higuchi, ``{Symmetric Tensor Spherical Harmonics on the $N$ Sphere and Their
  Application to the De Sitter Group SO($N$,1)},''
  \href{http://dx.doi.org/10.1063/1.527513}{{\em J. Math. Phys.} {\bfseries 28}
  (1987) 1553}. [Erratum: J.Math.Phys. 43, 6385 (2002)].

\end{thebibliography}\endgroup



\end{document}